\documentclass[fleqn,usenatbib]{mnras}

\usepackage{newtxtext,newtxmath}

\usepackage[T1]{fontenc}
\usepackage{ae,aecompl}

\usepackage{graphicx}	
\usepackage{amsmath}	
\usepackage{textgreek}
\usepackage{bm}

\newcommand*{\dint}{\, \mathrm{d}}
\newcommand*{\medusa}{{\sc{MEDUSA}} }

\title[Minkowski Functionals from Delaunay tessellation]{{\sc MEDUSA}: Minkowski functionals estimated from Delaunay tessellations of the three-dimensional large-scale structure}

\author[M. Lippich \& A. G. S\'anchez]{
 Martha Lippich\thanks{E-mail: mlippich@mpe.mpg.de} and
Ariel G. S\'anchez
\\
Max-Planck-Institut f\"ur extraterrestrische Physik, Postfach 1312, Giessenbachstr., 85741 Garching, Germany
}

\date{Submitted to MNRAS}

\pubyear{2020}

\begin{document}
\label{firstpage}
\pagerange{\pageref{firstpage}--\pageref{lastpage}}
\maketitle

\begin{abstract}
Minkowski functionals (MFs) are a set of statistics that characterise the geometry and topology of the 
cosmic density field and contain complementary information to the standard two-point analyses. We 
present {\sc MEDUSA}, an implementation of an accurate method for estimating the MFs of three-dimensional 
point distributions. These estimates are inferred from triangulated 
isodensity surfaces that are constructed from the Delaunay tessellation of the input point sample. 
Contrary to previous methods, \medusa can account for periodic boundary conditions, which is crucial 
for the analysis of N-body simulations. We validate our code against several test samples with 
known MFs, including Gaussian 
random fields with a \textLambda CDM power spectrum, and find excellent agreement with the 
theory predictions. We use \medusa to measure the MFs of synthetic galaxy catalogues constructed 
from N-body simulations. Our results show clearly non-Gaussian signatures that arise from the 
non-linear gravitational evolution of the density field. 
We find that, although redshift-space distortions significantly change our MFs estimates, 
their impact is considerably reduced if these measurements are expressed as a function of the volume-filling fraction. We also show that the effect of Alcock-Paczynski  (AP) distortions on the MFs can be 
described by scaling them with different powers of the isotropic AP parameter $q$ defined in terms 
of the volume-averaged distance $D_{\rm V}(z)$.
Thus the MFs estimates by MEDUSA are useful probes of non-linearities in the density field, and the expansion 
and growth of structure histories of the Universe.
\end{abstract}




\section{Introduction}

The analysis of the large-scale distribution of galaxies by means of two-point statistics, 
such as the correlation function or the power spectrum, has played a key role in 
establishing the current $\Lambda$CDM cosmological paradigm
\citep{Davis1983, Efstathiou2002,Tegmark2004,Cole2005,Eisenstein2005,Blake2011,Sanchez2006,
Sanchez2017,Alam2017,eBOSS2020}.
However, as the late time matter density field is not simply Gaussian distributed, 
two-point statistics cannot provide a complete description of the large-scale structure (LSS) of the Universe. 
Extending current cosmological analyses to higher order statistics is essential to 
exploit the full potential of upcoming high-precision surveys such as 
the dark energy spectroscopic instrument \citep[DESI,][]{desi_survey}, and
the ESA space mission {\it Euclid} \citep{Laureijs2011}. 

A full characterization of the density field would require measuring an infinite hierarchy of $N$-point 
correlation functions. 
Present-day surveys allow for accurate measurements of the three-point correlation function and 
the bispectrum \citep{Marin2013, Gil-Marin2015, 2017MNRAS.465.1757G, 2017MNRAS.468.1070S,
2017MNRAS.469.1738S,2018MNRAS.478.4500P}. 
These analyses are challenging due to the complexity associated with the measurement of all 
possible combinations of triplets, their corresponding theoretical modelling, and the estimation of accurate 
covariance matrices. The analysis of higher order  $N$-point functions is at the moment infeasible.

Additional statistics that encode compressed higher-order information have been proposed as 
alternatives to complement the standard two-point analyses. 
Among others, these include measurements of count in cells  
\citep[e.g.][]{1991ApJ...369..273D,1993ppc..book.....P,2020MNRAS.498L.125R,2020MNRAS.495.4006U}, 
the void probability function \citep[e.g.][]{1979MNRAS.186..145W, 2020arXiv201103556P}, 
and the genus \citep[e.g.][]{1986ApJ...306..341G, Weinberg1987, 2020ApJ...896..145A}.
In this work, we focus on the full set of Minkowski functionals (MFs), 
introduced to LSS studies by \citet{1994A&A...288..697M} to describe the geometry and 
topology of the cosmic density field. In three dimensions, there are four 
MFs: surface area, volume, curvature and the Euler characteristic (or the genus). 
For Gaussian density fields, the MFs of isodensity surfaces follow known analytical predictions 
that are sensitive to the power spectrum of the sample 
\citep{1990Tomita, 1997ApJ...482L...1S,2003ApJ...584....1M}. The 
non-linear gravitational evolution of the density field leads to deviations from Gaussianity, which 
manifest most notably in the asymmetry of the genus curve \citep{1994ApJ...434L..43M}, but also 
affect the other MFs. 
New theoretical predictions for the MFs of weakly non-Gaussian fields
\citep{2009PhRvD..80h1301P,2010PhRvD..81h3505M,2012PhRvD..85b3011G,2020arXiv201104954M,2020arXiv201200203M}, 
even accounting for redshift-space distortions (RSD) \citep{2013MNRAS.435..531C}, 
make it possible to access further information encompassing higher-order correlations, as well as the 
growth-rate of cosmic structure.
Additionally, other aspects of the evolution of density fluctuations can be explored using MFs, 
such as the subtle effect of massive neutrinos on the  morphology of the LSS \citep{2020arXiv200208846L}.

Although in this work we focus on the extraction of the MFs of the three-dimensional galaxy density field, 
there are several additional applications of these statistics in cosmological data analyses. 
Two-dimensional MFs have been extensively used to analyse non-Gaussianities in the CMB 
\citep[e.g.][]{1998MNRAS.297..355S,2006ApJ...653...11H,2017CQGra..34i4002B,2020A&A...641A...7P}. 
Weak lensing shear fields and convergence maps have also been investigated by means of MFs 
\citep[e.g.][]{2013PhRvD..88l3002P,2018PhRvD..98j3507S,2020arXiv201005669M,2020A&A...633A..71P}. 
Other fields of application are 21\,cm and reionization studies, for which MFs represent useful 
morphological descriptors \citep[e.g.][]{2015ApJ...814....6W,2017MNRAS.465..394Y,2019ApJ...885...23C}.

There are two main approaches to estimate MFs of the three-dimensional galaxy distribution. 
One method is to use germ-grain models that construct the MFs from intersecting 
spheres inflated around the input point sample 
\citep[see, e.g.,][for a comprehensive overview]{1994A&A...288..697M, 1996dmu..conf..281S,Kerscher_lecture}. The 
most useful aspect of this approach is that the resulting MFs can be directly expressed as sums over integrals of the 
$N$-point correlation functions \citep{1999MNRAS.309.1007S,2014MNRAS.443..241W}. The other popular approach 
is to estimate isodensity MFs, which are measured in terms of excursion sets. Isodensity MFs are more 
directly linked to the underlying density field than the ones from  germ-grain models, and have the 
advantage that they can be modelled with the previously mentioned theory predictions. Different 
methods to compute isodensity MFs from a galaxy distribution have been proposed, the most common 
ones are Koenderink invariants from differential geometry and Crofton's intersection formula 
from integral geometry \citep{1997ApJ...482L...1S,1999ApJ...526..568S}. 

MFs have been used to analyse LSS data for almost three decades, leading to interesting 
cosmological results. The germ-grain MFs have been used to study several 
galaxy and galaxy cluster catalogues 
\citep[e.g.][]{1994A&A...288..697M,1997MNRAS.284...73K,1998A&A...333....1K,2001A&A...377....1K,2014MNRAS.443..241W}. 
The recent work of \citet{2017MNRAS.467.3361W} studied the germ-grain MFs of the northern 
CMASS galaxy sample of SDSS BOSS DR12 and showed that they include contributions up to the 
sixth-point correlation function. Isodensity MFs have also been measured from LSS observations. 
Most analyses focused on the genus statistic only  
\citep[e.g.][]{2005ApJ...633...11P,2009ApJ...695L..45G,2009MNRAS.394..454J,2010ApJS..190..181C,2010ApJ...722..812Z}. 
Measurements of the four MFs based on Crofton's formula have been obtained from 
a preliminary SDSS sample by \citet{2003PASJ...55..911H} 
and the WiggleZ Dark Energy Survey \citep{Drinkwater2010}
by \citet{2014MNRAS.437.2488B}.

An alternative approach for the estimation of isodensity MFs is to compute these statistics on  triangulated 
isodensity surfaces constructed from the underlying density field. Although  this approach 
follows very closely the geometry of the isodensity surfaces, it has so far received little 
attention. \citet{2003MNRAS.343...22S} developed a code to construct triangulated surfaces from 
fixed lattice cubes and \citet{2004MNRAS.354..332S} used it to analyse N-body simulations with different 
cosmologies. \citet{2004BAAA...47..377Y} and \citet{ISVD.2010.33A} proposed to define the triangulated 
surface directly from a Delaunay tessellation of the galaxy distribution instead of using a regular grid.

This paper presents a new implementation of an algorithm to construct triangulated isodensity surfaces 
based on the Delaunay tessellation and to estimate their corresponding MFs. Our code, {\sc MEDUSA}, correctly 
accounts for periodic boundary conditions, which is crucial for the analysis of density fields from N-body 
simulations and the correct comparison of the measurements against theory predictions. 
After  validating \medusa thoroughly using a series of test samples for which the MFs can be theoretically 
predicted, we apply it to the analysis of synthetic galaxy catalogues based on the Minerva simulations 
\citep{2016MNRAS.457.1577G, Lippich2019}. We focus on three main issues of great importance 
for the analysis of the MFs of triangulated surfaces inferred from real galaxy surveys: 
non-Gaussian features due to non-linear gravitational evolution, redshift-space distortions (RSD), 
and Alcock-Paczynski (AP) distortions.

Our paper is structured as follows. In Section\,\ref{sec:mf_theo} we give a brief overview of MFs and describe 
the basic algorithm to measure them implemented in {\sc MEDUSA}. In Section\,\ref{sec:test_samples} we test 
\medusa by applying it to samples with known theoretical predictions for their MFs, including spherical, 
ellipsoidal and toroidal density distributions, as well as a Gaussian density field. 
In Section\,\ref{sec:densest} we discuss how we reconstruct the underlying density field from 
discrete point samples. In Section\,\ref{sec:mf_hod} we apply \medusa to the Minerva HOD galaxy 
samples in real and redshift space, and assess the impact of AP distortions. 
We present our conclusions in Section\, \ref{sec:conclusions}.

\section{Extracting Minkowski Functionals from a Delaunay tessellation}
\label{sec:mf_theo}

\subsection{Minkowski Functionals}

Minkowski Functionals (MFs) map the geometry and topology of an $N$-dimensional manifold into
a set of $N+1$ scalars. In our case, we consider as manifolds the 
excursion sets of the three-dimensional density field obtained by 
applying a given density threshold, $\rho_{\rm th}$. Points 
for which $\rho(\bm{r}) > \rho_{\rm th}$ are considered to be inside the isodensity surfaces. 
For this three-dimensional manifold there are four MFs:
\begin{enumerate}
    \item the \textit{surface area $S$},
    \item the \textit{volume $V$} enclosed by the surface,
    \item the \textit{integrated mean curvature $C$} of the surface,
    \begin{equation}
       C =  \frac{1}{2}\oint\limits_S \left(\frac{1}{R_1}+\frac{1}{R_2}\right)\dint S ,
    \end{equation}
    where $R_1$ and $R_2$ are the principal radii of curvature at a given point on the surface,
    \item the \textit{integrated Gaussian curvature} of the surface,
        \begin{equation}
       \chi =  \frac{1}{2\pi}\oint\limits_S \left(\frac{1}{R_1 R_2}\right)\dint S,
    \end{equation} also known as \textit{Euler-characteristic}.
\end{enumerate}
As can be seen from these equations, the MFs provide direct information about the geometry of the isodensity surface. In addition, the Euler characteristic also contains topological information, since it is related to the genus of the surface, $G = 1 - \chi/2$.
The genus is a fundamental quantity in topology and corresponds to the number of holes of the surface, or more precisely
\begin{equation*}
G = 1 + \text{number of holes} - \text{number of isolated regions},
\end{equation*}
where a hole is what is commonly referred to as tunnel in large-scale structure.
For example, the genus of a sphere is $G = 0$, the genus of a torus with one handle is $G = 1$ and the genus of 
an eyeglasses frame is $G=2$. Closed multiply-connected surfaces always have $G>0$. 
Furthermore, the MFs have some mathematical properties that are useful for large-scale structure analyses, 
in particular that they are additive and translational invariant.

It is common to work in terms of Minkowski functional densities, which are obtained by rescaling the global MFs
by the total volume considered, $V_{\rm tot}$. The rescaled volume functional is often referred to as the
volume-filling fraction
\begin{equation}
f_V = V/V_{\rm tot}. 
\end{equation}
We will denote the surface, curvature, and genus densities by $s$, $c$, and $g$, respectively. 
 
\subsection{The MEDUSA code}
\label{sec:medusa-code}
In real or synthetic galaxy catalogues in general we do not know the underlying continuous density field. 
Instead, we have to estimate the isodensity MFs from a discrete three-dimensional point distribution. 
For this, we need three main ingredients
\begin{enumerate}
    \item an estimate of the density at each point of the distribution,
    \item a fast and accurate extraction of the isodensity surfaces based on the point distribution 
    for any given density threshold,
    \item the computation of the MFs of the resulting isodensity surface.
\end{enumerate}
These three steps are implemented into our code \medusa (Minkowski functionals 
Estimated from DelaUnay teSsellAtion). There are several approaches to estimate
densities based on a discrete set of points. However, the second and third steps only require a set of points 
with known densities as an input and are independent of the particular method used to obtain such values. 
In the following two subsections, we describe each of the two steps (ii) and (iii) in detail and 
come to point (i) in Section\,\ref{sec:densest} after testing \medusa extensively on 
point distributions with known densities.

\subsubsection{Extraction of isodensity surfaces} 
\label{sec:delaunay}
\begin{figure}
 \centering
 \includegraphics[width=0.99\columnwidth]{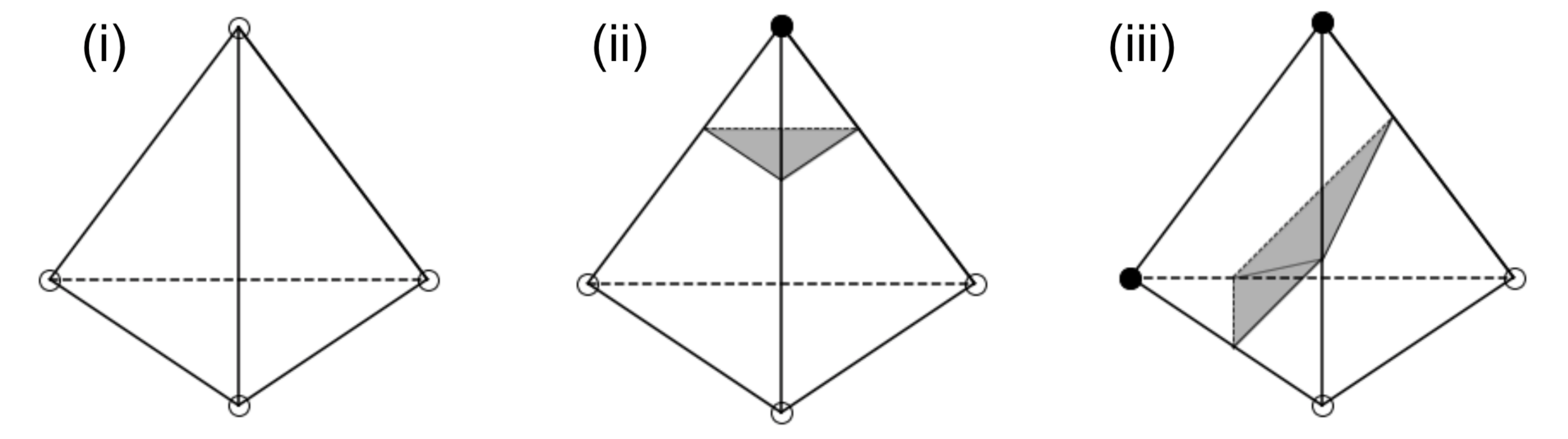}
  \caption{The three different tetrahedron configurations considered in {\sc{MEDUSA}}: (i) all vertices are the same, 
  either underdense or overdense compared to the density threshold $\rho_{\mathrm{th}}$, here shown as 
  empty circles, (ii) one vertex is different to the others, here shown as filled circle, (iii) two vertices are 
  underdense and the other two are overdense. The different fillings of the circle indicate the different densities. 
  The grey area shows the intersection of the triangulated isodensity surface with the tetrahedron.}
  \label{fig:tetra}
\end{figure} 

The crucial step for the estimation of the MFs is the extraction of the isodensity surfaces at a 
desired threshold from a set of points. For \medusa we chose a similar approach 
to \citet{2003MNRAS.343...22S}, \citet{2004BAAA...47..377Y} and \citet{ISVD.2010.33A}, and 
compute a triangulated isodensity surface directly from a three-dimensional point distribution. 
We extend these previous approaches by also including a recipe to account for periodic boundary 
conditions. Analogously to \citet{2004BAAA...47..377Y} and \citet{ISVD.2010.33A}, we perform 
a Delanauy tessellation on the 
three-dimensional point distribution, which we use as the basis for the interpolation 
of the density field. This approach is simpler than the regular grid used by 
\citet{2003MNRAS.343...22S} and automatically provides us with higher resolution in the 
regions where the density is higher.
In the case of point distributions from boxes with periodic boundary conditions, we add 
buffer zones around the box that replicate the particles from the opposite sides.
\medusa assigns a flag to each tetrahedron resulting from the Delaunay tessellation. This flag depends 
on how many particles of the tetrahedron are inside the box and, if the tetrahedron lies (partially) 
outside the box, on its position. The following cases need to be considered:
\begin{enumerate}
\item Tetrahedra that lie completely inside the box.
\item Tetrahedra that are partially outside the box and cross one face of the box far from its edges. Each of these tetrahedra 
has one copy at the opposite side of the box.
\item Tetrahedra that are partially outside and lie close to the edges, but far from the corners of the box. Each of these tetrahedra has three copies at the three opposite edges of the box.
\item Tetrahedra which are partially outside and  lie close to the corners of the box. Each of these tetrahedra has seven copies at the other seven corners of the box. 
\item For tetrahedra that are located at the corners of the box there is the special case that the vertices of the tetrahedron are all outside, but the tetrahedron is still partially inside the box.
\item Tetrahedra close to the edges or corners that lie completely outside the box and are copies of 
tetrahedra that are completely inside the box, but are neighbours of tetrahedra that are partially inside.
\item Tetrahedra that are completely outside the box and do not belong to the previous case (vi). 
They are also copies of tetrahedra that are completely inside the box.
\end{enumerate}
The  tetrahedra that belong to the last category (vii) can be discarded. All other tetrahedra are assigned a flag that takes into 
account to which category they belong and at which side of the box they are located, in order to prevent double counting. 
These flags are used in the estimation of the MFs as described in Section\,\ref{sec:mf_est}.
 
Additionally, all particles are considered as ``overdense" or ``underdense" depending on whether their corresponding densities 
are larger than the density threshold $\rho_{\mathrm{th}}$ being considered or not. 
Given this classification, there are only three different types of tetrahedron configurations:
\begin{enumerate}
    \item All vertices of the tetrahedron are either overdense or underdense. 
    \item One vertex is different to the other three vertices.
    \item Two vertices are overdense and the other two are underdense.
\end{enumerate}
These three cases are illustrated in Fig.\,\ref{fig:tetra}, where vertices with the same density property, i.e. underdense 
or overdense, are shown with circles with the same filling. 
The isodensity surface will only intersect tetrahedra of the last two types. 
This intersection will occur at the edges between particles with different density properties. 
The intersection points of the surface with the tetrahedron edges correspond to the points where the density matches 
the threshold $\rho_{\mathrm{th}}$, which are obtained by linearly interpolating the densities of the two 
corresponding particles. 
This is equivalent to assuming a constant density gradient within the tetrahedron. 
For case (ii), where one particle is different to the others, we obtain an intersection triangle. For case (iii), where two 
particles have the same density property, we obtain four points of intersection on the edges and the 
resulting surface can be decomposed into two triangles. 
Following this approach, \medusa computes the intersection triangles for 
all tetrahedra where at least one vertex is different to the others. 
These are 12 configurations less to take into account than for cubic lattice 
intersections as in \citet{2003MNRAS.343...22S}, which makes this step significantly simpler. 
Once all tetrahedra of types (ii) and (iii) have been considered, we obtain a 
triangulated surface representing the isodensity contour corresponding to 
$\rho_{\mathrm{th}}$.

\subsubsection{Minkowski Functionals of a triangulated surface}
\label{sec:mf_est}
Since the MFs are additive, the global MFs of the density distribution can be obtained by summing over 
the MFs of the isodensity surfaces enclosing the individual excursion sets. 
As described in \citet{2003MNRAS.343...22S}, the MFs of a triangulated 
surface can be computed in a straightforward way: 
\begin{enumerate}
 \item The surface area $S$ of the triangulated surface is given by the sum over the areas of all triangles of the surface,
 \begin{equation}
    S = \sum _{i=1} ^{N_\mathrm{t}} S_i,
    \label{eq:surf}
 \end{equation}
   where $N_{\mathrm{t}}$ is the total number of triangles contributing to the surface.
 \item The volume $V$ is the sum over the volumes of all fully enclosed tetrahedra, denoted with ${\mathrm{T}}$, and 
 the fraction of the volumes of the intersected tetrahedra that lie within the surface, denoted with ${\mathrm{S}}$,
  \begin{equation}
   V = \sum _{i=1} ^{N_\mathrm{T}} V_i + \sum _{j=1} ^{N_\mathrm{S}} V_j.
   \label{eq:vol}
  \end{equation}
  If only one vertex is overdense or underdense, corresponding to case (ii) in Fig.~\ref{fig:tetra}, the volume of the tetrahedron 
  defined by this point and the triangle of the isodensity surface as a base has to be added or subtracted, respectively. 
  If the tetrahedron contains two overdense vertices, as in case (iii) of Fig.~\ref{fig:tetra}, the contributing volume can be split into 
  three tetrahedra. 
  \item The integrated mean curvature $C$ is obtained by summing over the edges of all adjacent triangles $i$ and $j$,
  \begin{equation}
  C = \frac{1}{2}\sum _{i,j}\ell_{ij}\phi_{ij}\epsilon
  \label{eq:curv}
  \end{equation}
  where $\ell_{ij}$ is the length of the common edge, $\phi_{ij}$ is the angle between the normals, $\hat{n}_{i}$ and $\hat{n}_{j}$, of the two triangles,
  \begin{equation}
    \cos \phi_{ij} = \hat{n}_{i} \cdot \hat{n}_{j},
  \end{equation}
and the value of $\epsilon$ distinguishes the cases in which the surface is locally convex, indicated by the value 
$\epsilon =  1$, or locally concave, in which case $\epsilon = - 1$.
  \item The Euler characteristic $\chi$ of a triangulated surface can be determined by
\begin{equation}
    \chi = N_{\mathrm{t}} - N_{\mathrm{e}} + N_{\mathrm{v}},
    \label{eq:euler}
\end{equation}
where $N_{\mathrm{t}}$, $N_{\mathrm{e}}$ and $N_{\mathrm{v}}$ are the total number of triangles, triangle edges and 
triangle vertices contained in the surface.
\end{enumerate}

As mentioned in Section\,\ref{sec:delaunay}, \medusa assigns a flag to every tetrahedron depending on 
its position in the box and/or the buffer zone.  
From the tetrahedra that are partially inside the box, and hence are repeated on its other sides, only those 
that are closer to the origin (0,0,0) are taken into account in the sums of equations~(\ref{eq:surf})-(\ref{eq:vol}), while 
all other copies are discarded. 
The flags that \medusa assigns to each tetrahedron ensure that all triangle edges and vertices from tetrahedra that 
are partially inside the box are taken into account in the sums of equations~(\ref{eq:curv})-(\ref{eq:euler}), 
and that their contribution is counted only once. 

\begin{figure}
 \centering
 \includegraphics[width=\columnwidth]{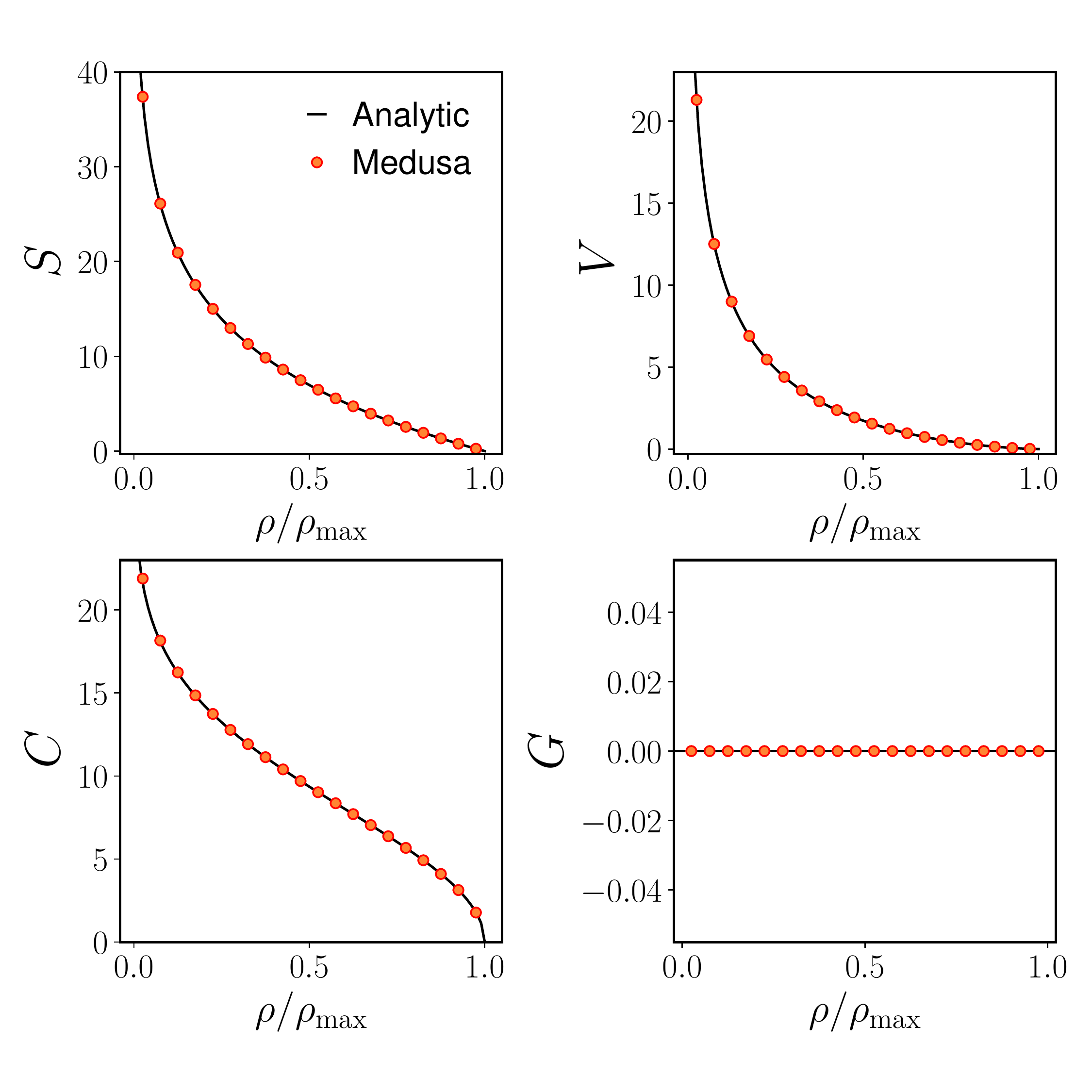}
  \caption{Minkowski Functionals inferred from a set of points following a spherically-symmetric 
  Gaussian distribution as a function of  the normalized density threshold $\rho/\rho_{\mathrm{max}}$. 
  The red circles correspond to the measurements from \medusa using 20 equispaced density 
  thresholds, the black lines show the analytical predictions.}
  \label{fig:sphere}
\end{figure}

\section{Results for test models}
\label{sec:test_samples}
In this section we test the performance of \medusa by measuring the MFs of point distributions following 
known density profiles. 

\begin{figure}
 \centering
 \includegraphics[width=0.95\columnwidth]{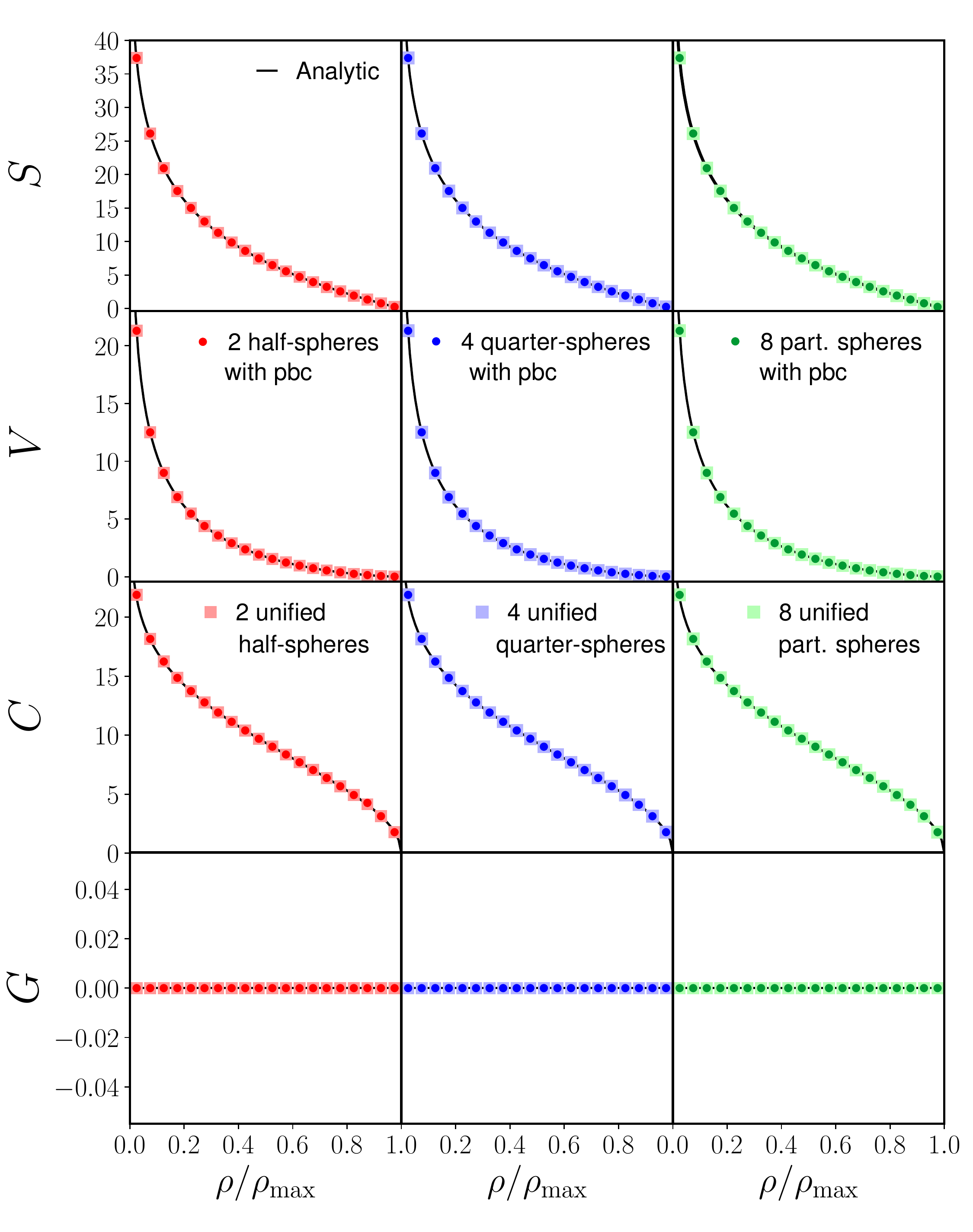}
\caption{Minkowski Functionals of spherical density distributions where the spheres were cut into 
two half-spheres located at two opposite faces of a cubic box (red), four quarter-spheres located 
at the centres of four opposite edges of the box (blue), and eight partial spheres located at 
each corner of the box (green), measured with periodic boundary conditions. The 
measurements agree perfectly with the ones obtained from the corresponding unified spheres 
and their analytical predictions. All MFs are measured as a function of threshold 
$\rho/\rho_{\mathrm{max}}$.}
  \label{fig:part_sphere_248}
\end{figure} 

\subsection{Spherical density distribution}
\label{sec:sph_dens}
As a first test sample we consider a distribution of points following a spherically-symmetric 
Gaussian density profile given by
\begin{equation}
    \rho(\bm{r}) = \rho_{\mathrm{max}} \exp\left(-\frac{r^2}{2\sigma^2}\right),
    \label{eq:sphere_an}
\end{equation}
where $r = \left|\bm{r}\right|$. We generated a set of points following this density profile with 
$3.5\times 10^5$ particles, $\sigma = 0.6$ and a maximum radius, $r_{\mathrm{max}} = 4.0$. 
The density at each point was obtained by evaluating the true density profile of 
equation~\eqref{eq:sphere_an} at the corresponding location. 
We used \medusa to measure the MFs of 20 equispaced density thresholds 
from $\rho/\rho_{\mathrm{max}} = 0.0$ to $1.0$.
The analytical MFs for a sphere are:
\begin{enumerate}
    \item $\begin{aligned}  S(\rho_{\rm th}) &= 4\pi r(\rho_{\rm th})^2 \end{aligned}$
    \item $\begin{aligned} V(\rho_{\rm th}) &= \frac{4}{3} \pi r(\rho_{\rm th})^3 \end{aligned}$
    \item $\begin{aligned} C(\rho_{\rm th}) &= 4 \pi r(\rho_{\rm th}) \end{aligned}$
    \item $\begin{aligned}\chi(\rho_{\rm th}) = 2 \text{ and hence }G = 0 \end{aligned}$ 
\end{enumerate}
The radius corresponding to a given density threshold, $r(\rho_{\rm th})$, can be 
obtained by inverting equation\,(\ref{eq:sphere_an}). 
A lower density threshold corresponds to a larger radius of the spherical isodensity surface. 
Fig.~\ref{fig:sphere} shows that the measured MFs are in good agreement with the analytical predictions. 
The overall agreement of the measurements of the first three MFs with the corresponding predictions 
is significantly better than 1\%. The measured genus is always zero.

In order to test the implementation of periodic boundary conditions, we generated sets of 
points following the same spherically symmetric density profile of equation~(\ref{eq:sphere_an}) 
but where the density distributions were cut into two half-spheres located at two opposite faces 
of a cubic box, four quarter-spheres located at the center of four opposite edges of the box, 
and eight partial spheres located at each corner of the box. Without the implementation of 
periodic boundary conditions, the isodensity surface cannot be extracted correctly at the boundaries 
of the box, since tetrahedra cannot extend outside it.
Fig.\,\ref{fig:part_sphere_248} shows the agreement between the MFs measured from these 
three distributions taking into account periodic boundary conditions and the results 
obtained from the unified spherical distribution, for which no periodic boundary conditions are required. 
The agreement with the analytical predictions is also excellent. 
These results show that \medusa can correctly account for distributions with periodic boundary 
conditions. In particular, an error in the implementation of the periodic boundary conditions leading 
to a single incorrectly counted triangle, triangle vertex or triangle edge would result in values 
of genus $G\ne0$.

\begin{figure}
 \centering
 \includegraphics[width=\columnwidth]{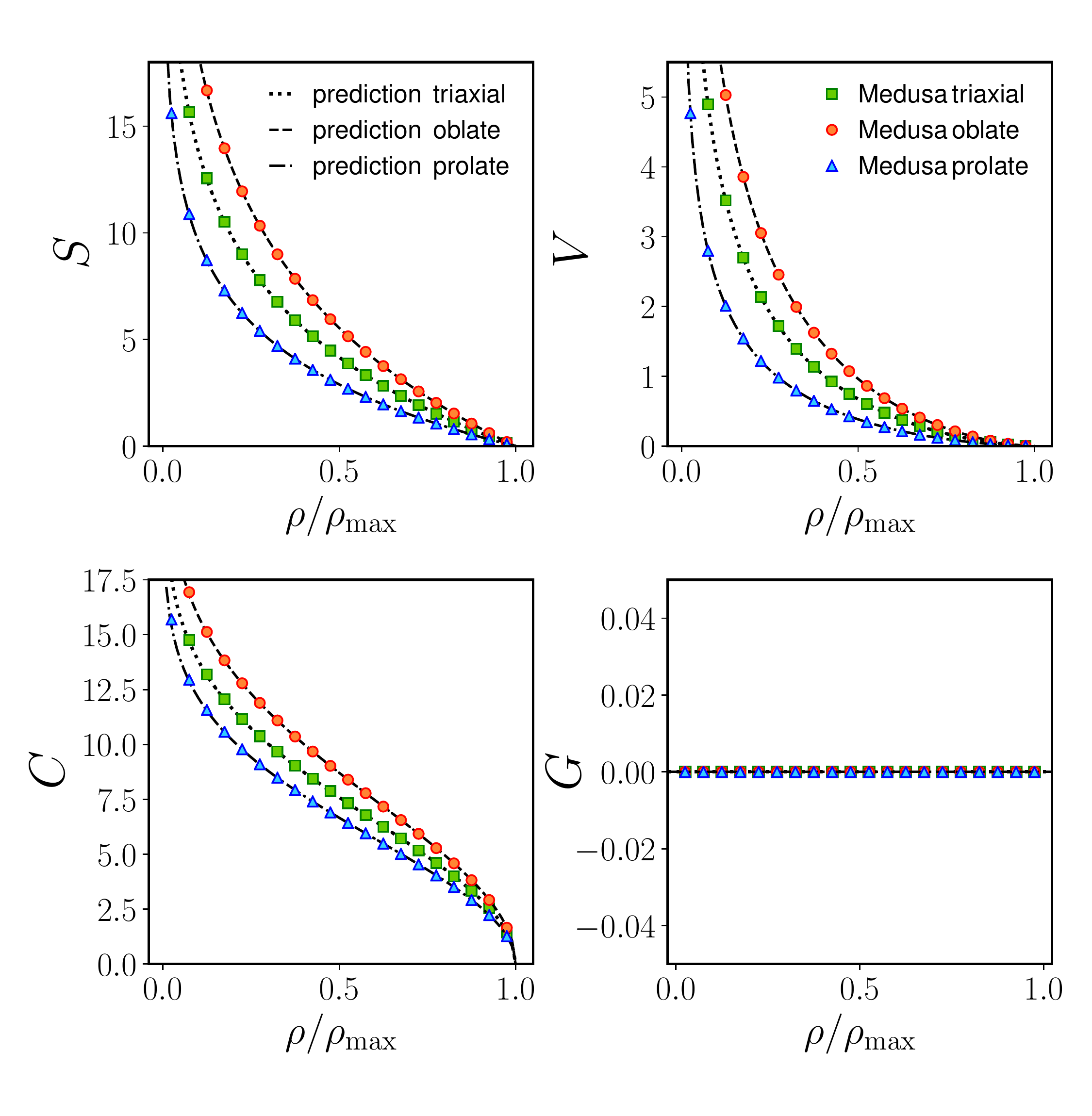}
  \caption{Minkowski Functionals for three different ellipsoidal point distributions, oblate, prolate, 
  and triaxial, defined by the density profile of equation~(\ref{eq:dens_ellipsoid}) 
  expressed as a function of the normalized density $\rho/\rho_{\rm max}$. The lines indicate the 
  corresponding theoretical predictions.}
  \label{fig:ellipsoid}
\end{figure}

\subsection{Ellipsoidal density distributions}

We now consider as test samples ellipsoidal density distributions given by,
\begin{equation}
    \rho(x,y,z) = \rho_{\mathrm{max}} \exp\left[-\left(\frac{x^2}{\sigma_a^2}+\frac{y^2}{\sigma_b^2}+\frac{z^2}{\sigma_c^2}\right)\right].
    \label{eq:dens_ellipsoid}
\end{equation}
We generated three different point distributions corresponding to oblate 
($\sigma_a = \sigma_b = 1.0$, $\sigma_c = 0.4$), 
prolate ($\sigma_a = \sigma_b = 0.4$, $\sigma_c = 1.0$), and 
triaxial ($\sigma_a = 0.4$, $\sigma_b = 0.7$, $\sigma_c = 1.0$) ellipsoids 
using the same number of points and density thresholds as for the spherical case.

For the oblate and the prolate case, we compared the measured Minkowski functionals against 
analytical predictions. For the triaxial case no analytical predictions are known for the surface and 
the curvature and therefore we computed numerical predictions. 
As for the case of  the spherical distributions, a lower density threshold corresponds to larger 
principal axes $a,b,c$ of the ellipsoid, while conserving the constant axes ratios for all density 
thresholds, such that $b = a\frac{\sigma_b}{\sigma_a}$ and $c = a\frac{\sigma_c}{\sigma_a}$. 
The analytical predictions for the MFs are given by:
\begin{enumerate}
    \item \begin{align}
        S_{\rm obl} &= 2\pi a\left(\rho_{\rm th}\right)^2 \left[ 1 + \frac{\sigma_c ^2} {\sigma_a\sqrt{\sigma_a^2-\sigma_c^2}} \mathrm{arctanh} \left(\sqrt{1-\frac{\sigma_c^2}{\sigma_a^2}} \right)\right] \\
        S_{\rm pro} &= 2\pi a\left(\rho_{\rm th}\right)^2 \left[ 1 + \frac{\sigma_c^2}{\sigma_a\sqrt{\sigma_c^2-\sigma_a^2}} \mathrm{arcsin} \left(\sqrt{1-\frac{\sigma_a^2}{\sigma_c^2}}\right)\right]
    \end{align}
    \item \begin{align}
     C_{\rm obl} &= 2\pi a\left(\rho_{\rm th}\right)\left[\frac{\sigma_c}{\sigma_a} + \frac{\sigma_a}{\sqrt{\sigma_c^2 - \sigma_a^2}}\mathrm{arccosh}\left(\frac{\sigma_c}{\sigma_a}\right)  \right] \\ 
     C_{\rm pro} &= 2\pi a\left(\rho_{\rm th}\right)\left[\frac{\sigma_c}{\sigma_a} + \frac{\sigma_a}{\sqrt{\sigma_a^2 - \sigma_c^2}}\mathrm{arccos}\left(\frac{\sigma_c}{\sigma_a}\right)  \right]
    \end{align}
    \item $ V = \frac{4}{3} \pi\, a\left(\rho_{\rm th}\right)^3 \frac{\sigma_b \sigma_c}{\sigma_a^2} $
    \item $\chi = 2$ and hence $G = 0$
\end{enumerate}
Fig.~\ref{fig:ellipsoid} shows that in all cases the measured MFs agree perfectly with the theoretical 
predictions.  As in the case for the spherical density distribution, the overall agreement of the 
measurements of the first three MFs with the corresponding predictions is better than 1\%. 
The measured genus is always exactly zero.

\begin{figure}
 \centering
 \includegraphics[width=0.7\columnwidth]{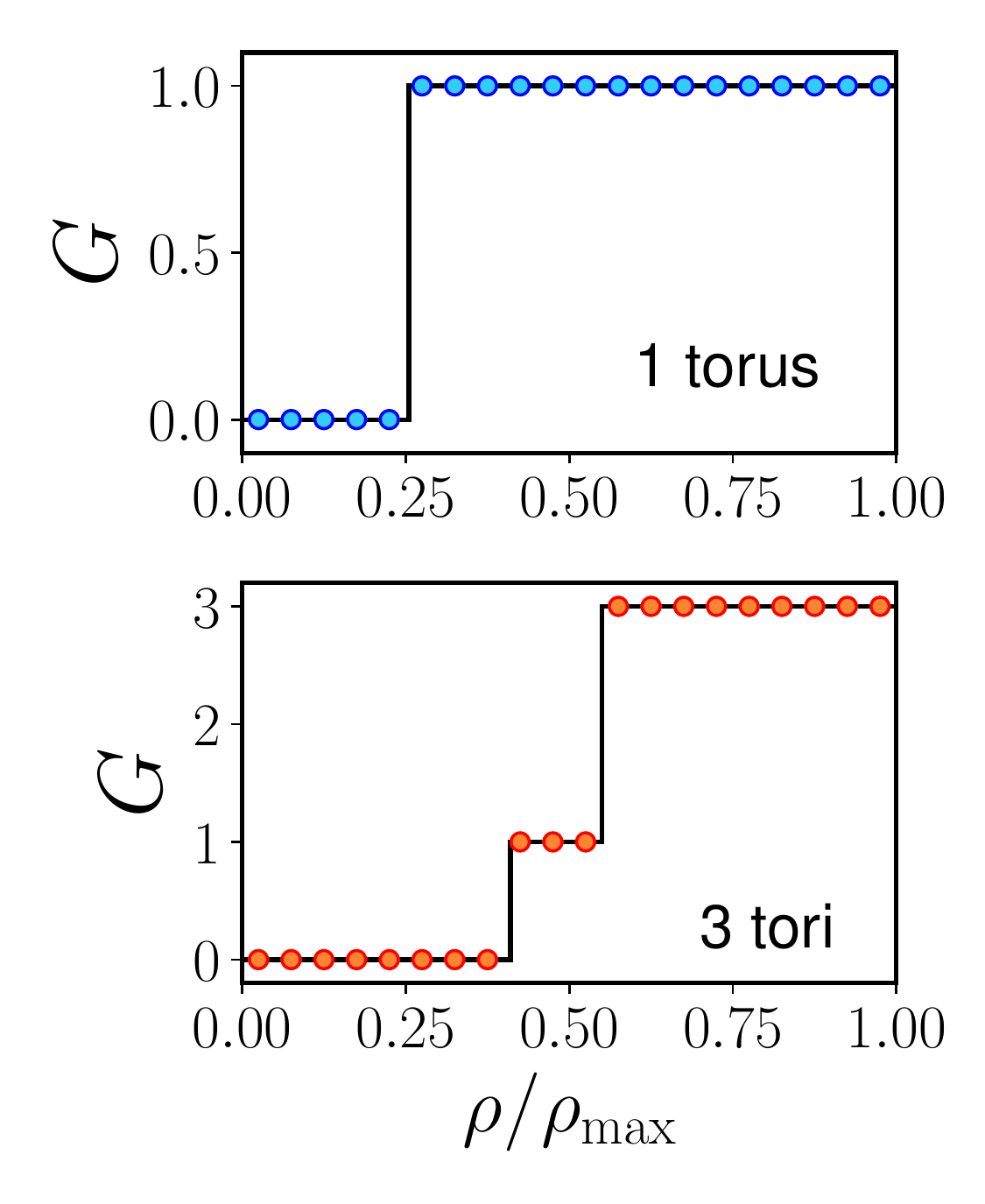}
  \caption{Genus measurements as a function of the normalized density $\rho/\rho_{\rm max}$ 
  of point distributions following profiles of one torus (upper panel) and  three overlapping 
  tori (lower panel). The solid lines show the theoretical predictions for each case.}
  \label{fig:torus}
\end{figure}

\begin{figure*}
 \centering
 \includegraphics[width=0.8\textwidth]{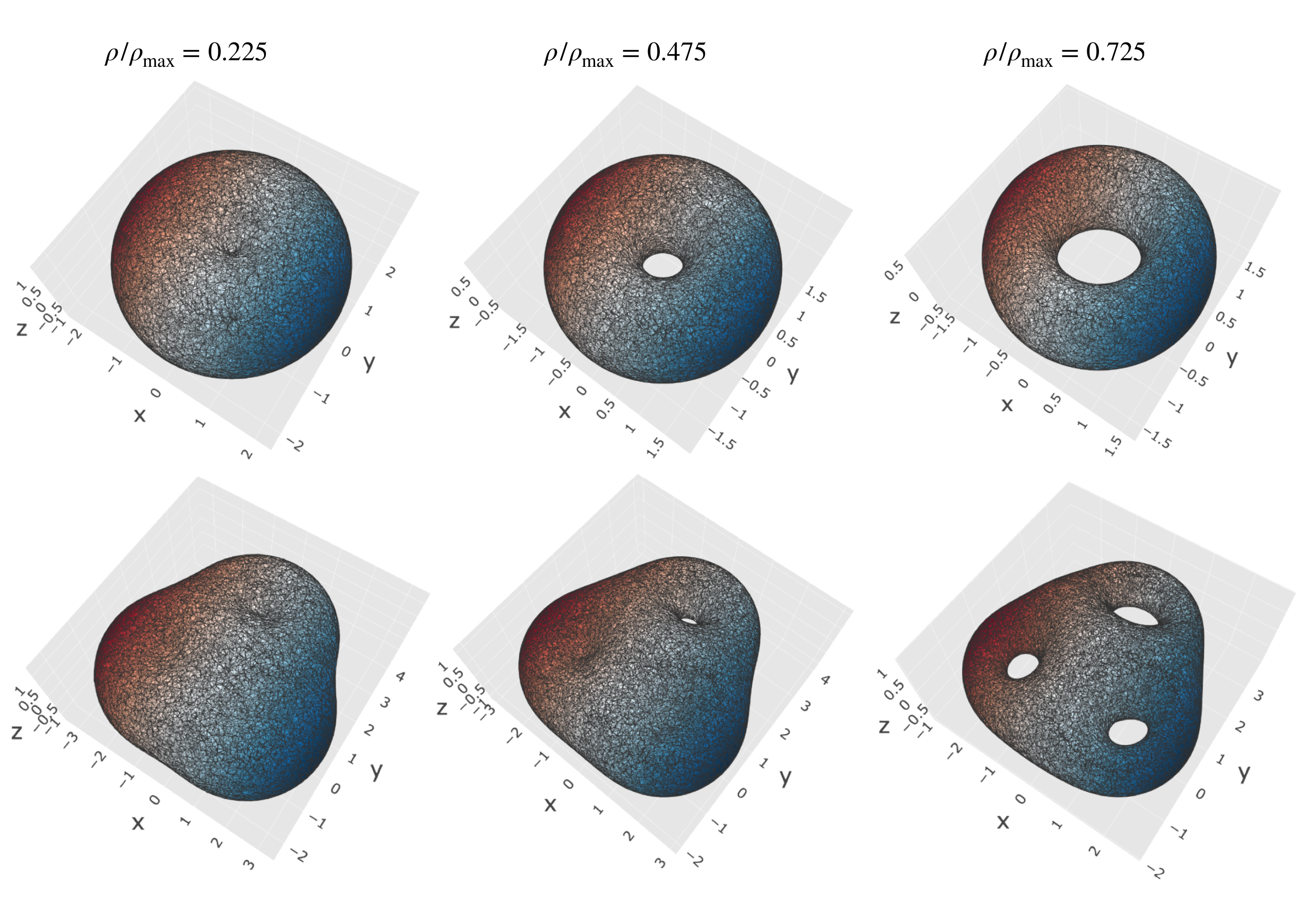}
  \caption{Isodensity surfaces for three different density thresholds $\rho/\rho_{\rm max} = 0.225$, $\rho/\rho_{\rm max}  = 0.475$  and  $\rho/\rho_{\rm max}  = 0.725$ for density distributions following profiles of one torus (upper panel) and  three overlapping tori (lower panel).}
  \label{fig:torus_surface}
\end{figure*}

\subsection{Toroidal density profiles}

The point distributions considered in the previous sections have isodensity surfaces without 
holes, and therefore their genus is zero for all density thresholds. In order to test the estimation 
of the Euler characteristic and the genus, we studied sets of points corresponding to one or 
more overlapping toroidal distributions.
For the case of one torus, the point distribution is generated following a density profile
\begin{equation}
    \rho(x,y,z) = \rho_{\rm max} \exp\left[- \frac{(R - \sqrt{x^2 + y^2})^2 + z ^2}{\sigma ^2}\right],
\end{equation}
where $R$ and $r$ are the major and minor radii of the torus, respectively, and 
$r^2 = \left(R - \sqrt{x^2 + y^2}\right)^2 + z^2$. The upper panel of
Fig.~\ref{fig:torus} shows the measured genus of such a density distribution, generated with 
$R=1.1$ and $\sigma = 0.9$. The upper panels of Fig.~\ref{fig:torus_surface} show the corresponding 
triangulated isodensity surfaces obtained by \medusa for $\rho/\rho_{\rm max}=0.225$, 
$\rho/\rho_{\rm max}=0.475$  and $\rho/\rho_{\rm max}=0.725$. For low density thresholds, no hole is visible in 
the isodensity surface and \medusa measures $G = 0$. For density thresholds 
$\rho/\rho_{\rm max} > 0.25$ a hole in the center of the isodensity surface becomes visible 
and the code correctly recovers $G = 1$.

We also considered a set of points corresponding to three overlapping tori with $R=1.3$ and 
$\sigma = 0.8$, and centred at $(1,0,0)$, $(-1,0,0)$ and $(0,2,0)$, respectively. The lower panel 
of Fig.~\ref{fig:torus} shows the genus measured from this density distribution, and the lower panels 
of Fig.~\ref{fig:torus_surface} show three characteristic isodensity surfaces at the same thresholds 
as before. As in the case of a single torus, for low density thresholds the isodensity surface contains 
no holes and the measurement of the genus is $G=0$. For a density threshold 
$0.4<\rho/\rho_{\rm max}<0.55$ the corresponding isodensity surface shows the hole of the torus whose 
center is furthest from the other two and hence the measured  genus is $G=1$. For a density 
threshold $\rho/\rho_{\rm max}>0.55$, the isodensity surface contains three holes and we recovered 
the correct value $G=3$.

\subsection{Effect of using particles tracing the density field}
\label{sec:random_points} 
\begin{figure}
 \centering
 \includegraphics[width=\columnwidth]{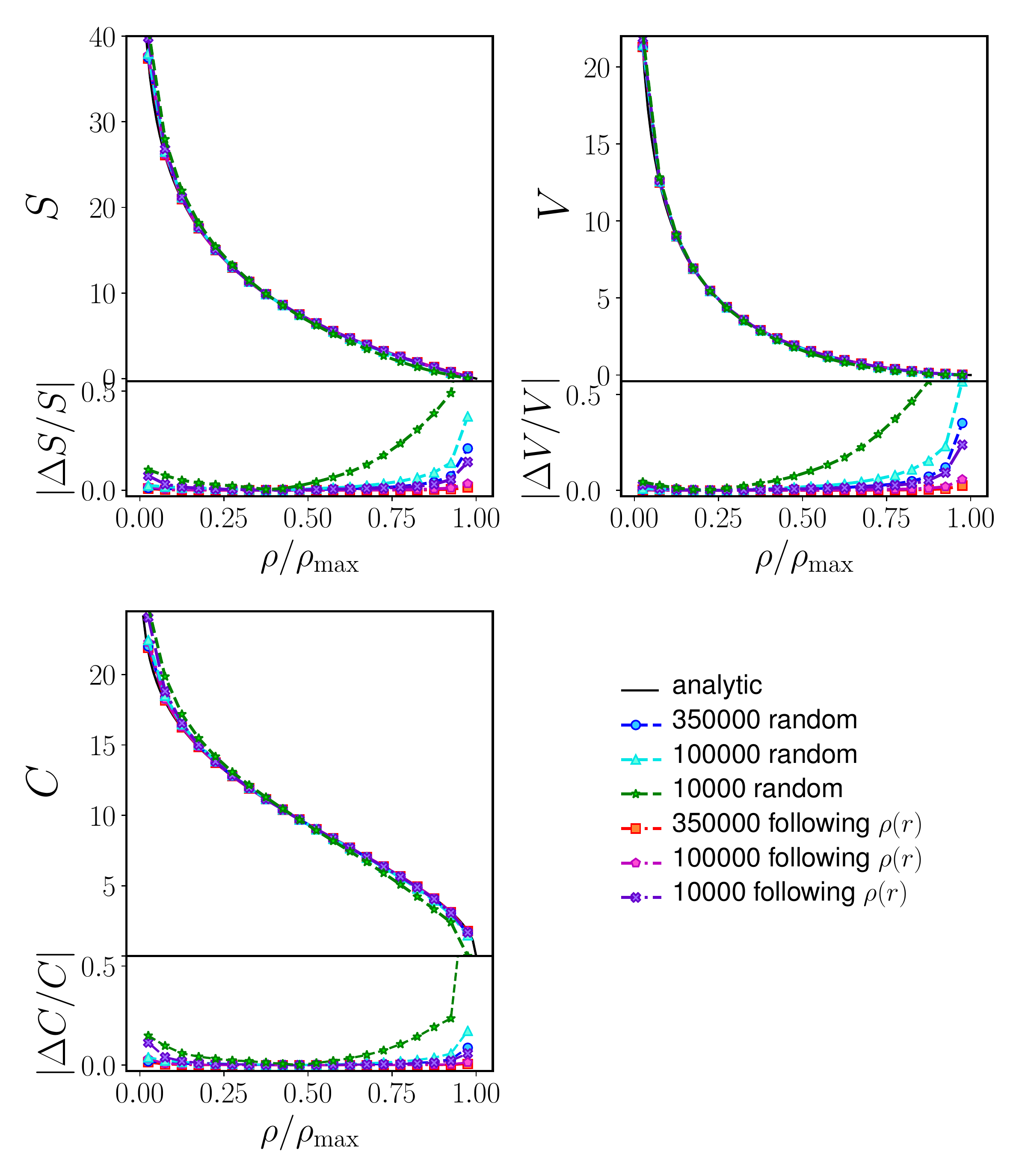}
  \caption{Minkowski Functionals for the spherical density distribution 
of Section~\ref{sec:sph_dens} obtained using samples with 
350\,000, 100\,000 and 10\,000 points that are placed following the density 
profile or randomly within the same volume. 
The genus is not shown as it is consistently zero in all cases.}
  \label{fig:sphere_ran}
\end{figure}

When estimating MFs, \medusa uses the values of the density field directly at the positions of the 
points in the sample being analysed. 
In the test samples considered in the previous sections, as in numerical simulations or real galaxy surveys, 
the points trace the underlying density field. Using their positions as the nodes to interpolate 
the density field, as opposed to, e.g. the vertices of a regular grid, has the advantage of automatically providing a 
higher resolution in high-density regions. Note however, that the procedure described in 
Section~\ref{sec:medusa-code} does not require the points used as the basis of the Delaunay tessellation 
to follow the density field.
As a test, Fig.~\ref{fig:sphere_ran} shows the MFs for the same spherical density distribution as in 
Section~\ref{sec:sph_dens} estimated using 
sets of particles of different size that are placed following the density distribution or randomly within 
the same volume. 
The computation of the genus is consistently zero for all considered cases and density thresholds. 
The remaining three MFs computed from the 100\,000 and 350\,000 particles tracing the density field 
agree with the analytical predictions at better than 1\% level. Even for the case of 10\,000 points the agreement 
between measurements and analytical predictions is better than 2\% on densities $0.1 < \rho/\rho_{\rm max} < 0.8$. 
For the case of the randomly distributed particles, we obtain a comparable precision only when using 350\,000 
particles. For smaller samples, the deviations from the analytical predictions become significantly larger, particularly 
for high density thresholds.  This comparison illustrates the advantage of using particles tracing the density field 
as the nodes of the Delaunay tessellation, which provides a better resolution on high-density regions and 
allows for a robust determination of all MFs even for sparse samples.

\subsection{Gaussian density field}

For a final test of {\sc{MEDUSA}}, we computed the MFs of a smoothed Gaussian random field (GRF), 
which have known analytical expressions that are sensitive to the power spectrum of the field, $P(k)$. 
This case also serves as an additional validation of the implementation of periodic 
boundary conditions in our code, as an incorrect  
treatment would lead to deviations from the analytical predictions. We generated 100 
realizations of a GRF with the same linear power spectrum as our Minerva simulations, which are 
described in Section\,\ref{sec:mf_hod} at redshift $z =0.57$ on a cubic grid with side length 
$L=737\,{\rm Mpc}$ and periodic boundary conditions.
The field, $f$, was smoothed with a Gaussian kernel 
\begin{equation}
W(x)= \frac{1}{(2\pi)^{3/2}R^3}\exp\left(-\frac{x^2}{2R^2}\right),
\label{eq:gaussian_kernel}
\end{equation}
with a smoothing scale $R = 20$ grid units ($=28.8\,$Mpc) and normalized by its standard deviation, 
$\nu=f/\sigma_0$, such that it has zero 
mean, $\langle \nu \rangle = 0$, and unit variance, $\langle \nu^2 \rangle = 1$. Since there are grid 
cells with negative values of $\nu$, it cannot be treated as a density field and sampled with points. 
Instead, we follow the approach tested in Section~\ref{sec:random_points} and use the values of $\nu$ 
at 200\,000 randomly placed points in each box. The resulting mean interparticle separation is 
approximately half of the smoothing length, and thus it should be possible to resolve the full structure 
of the smoothed density field. 

\begin{figure}
 \centering
 \includegraphics[width=\columnwidth]{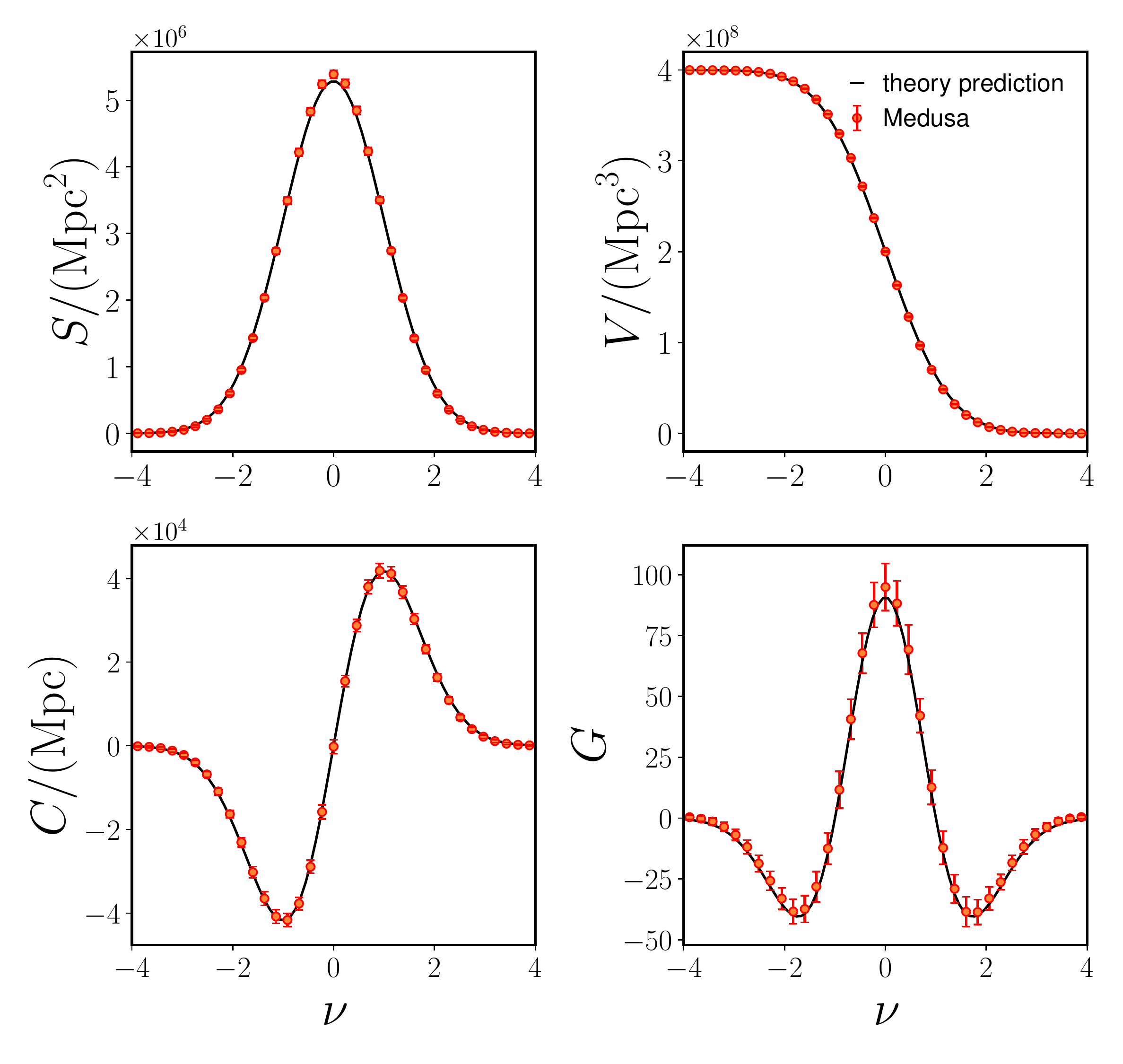}
  \caption{Mean MFs of 100 GRFs, generated with the same linear 
  $\Lambda$CDM power spectrum as the Minerva simulations in a cubic box with length $737\,{\rm Mpc}$ 
  and smoothed with a Gaussian kernel with $\lambda = 20$ grid units ($\lambda=28.8\,{\rm Mpc}$). We used 
  the densities at 200\,000 random points within the box. The red points show the mean values determined 
  using \medusa together with the standard deviation from 100 realizations. The theoretical 
  predictions are shown as black solid lines.}
  \label{fig:gauss}
\end{figure}

The theoretical predictions for the MFs 
of a smoothed GRF only depend on the parameter $\lambda_{\rm c}$, given by
 \begin{equation} 
 \lambda_{\rm c} = \sqrt{\frac{2\pi\xi(0)}{|\xi''(0)|}},
 \end{equation}
 which can be derived from the value of the correlation function $\xi(0)$ and its second  derivative 
 $\xi''(0)$ at zero separation. Both can be directly computed from the power spectrum 
 of the sample as
  \begin{align}
   \xi(0) &= \langle f^2 \rangle = \sigma_0 ^2, \\
   |\xi''(0)| &=  \langle |\nabla f|^2 \rangle = \sigma_1 ^2, \text{ and} \\
   \sigma_j ^2(R) &= \int \frac{k^ 2 \dint k}{2\pi^2}k^{2j}P(k)\hat{W}(kR)^2,
 \end{align}
 in which $\hat{W}(kR)$ is the Fourier transform of the filter kernel of equation~(\ref{eq:gaussian_kernel}). 
 For a GRF in three dimensions, the four MF densities are given by
 \citep{1990Tomita, 1997ApJ...482L...1S,2003ApJ...584....1M}: 
\begin{align}
        \label{eq:gauss-vol}
        f_V(\nu) &= \frac{1}{2} -  \frac{1}{2} \Phi\left(\frac{\nu}{\sqrt{2}}\right),\\
        \text{where } \Phi(x) &= \frac{2}{\sqrt{\pi}}\int _0 ^x \dint t \exp\left(-t^2\right) \text{ denotes the error function},\\
        \label{eq:gauss-surf}
        s(\nu) &= \frac{2}{\lambda _c} \sqrt{\frac{2}{\pi}} \exp{\left(-\frac{\nu^2}{2}\right)}, \\
        \label{eq:gauss-curv}
        c(\nu) &= \frac{\sqrt{2\pi}}{\lambda _c^2}\nu \exp{\left(-\frac{\nu^2}{2}\right)}, \\
        \label{eq:gauss-gen}
        g(\nu) &= \frac{1}{\lambda_c ^3 \sqrt{2\pi}}(1-\nu^2)\exp{\left(-\frac{\nu^2}{2}\right)}.
\end{align}
From these expressions, it can be seen that the amplitudes of three of the MFs of a GRF are sensitive 
to its underlying power spectrum.
Fig.\,\ref{fig:gauss} shows the mean MFs computed with \medusa from the 100 realizations of the 
GRF and their corresponding theoretical predictions, which are in good agreement. 
This shows that \medusa can accurately determine MFs of cosmological density 
fields and that the periodic boundary conditions are correctly implemented.

\section{Density estimation}
\label{sec:densest}

The procedure to extract isodensity surfaces and estimate MFs described in Section~\ref{sec:medusa-code}
requires as input the values of the density at each point of our discrete distribution.
In the test cases of Section~\ref{sec:test_samples}, we used the true values of the underlying density field
evaluated at the position of the points. When analysing N-body simulations or galaxy surveys, these 
densities need to be estimated from the point distribution itself.   
Here, we estimate densities by applying a Gaussian kernel with a fixed smoothing scale $\lambda$,
\begin{equation}
    W(r) = \frac{1}{A} \exp{\left(-\frac{r^2}{2\lambda^2}\right)},
    \label{eq:Gaussian_kernel}
\end{equation}
where $r$ represents the distance between the points. This kernel is truncated at a scale $r_{\rm cut}$
and the normalization $A$ is defined such that the volume integral of $W(r)$ up to this maximum scale 
is equal to one and hence the total mass is conserved. 
We tested this approach by applying it to the spherical Gaussian density distribution described in 
Section~\ref{sec:sph_dens} for which the true underlying density distribution is known.
We examined the impact of using different kernel smoothing lengths and truncation radii.
The true underlying density profile corresponding to each case can be obtained by convolving the Gaussian 
density field of equation~(\ref{eq:sphere_an}), which is truncated at $r_{\mathrm{max}} = 4$, with 
the kernel of equation~(\ref{eq:Gaussian_kernel}).

\begin{figure}
 \centering
 \includegraphics[width=0.8\columnwidth]{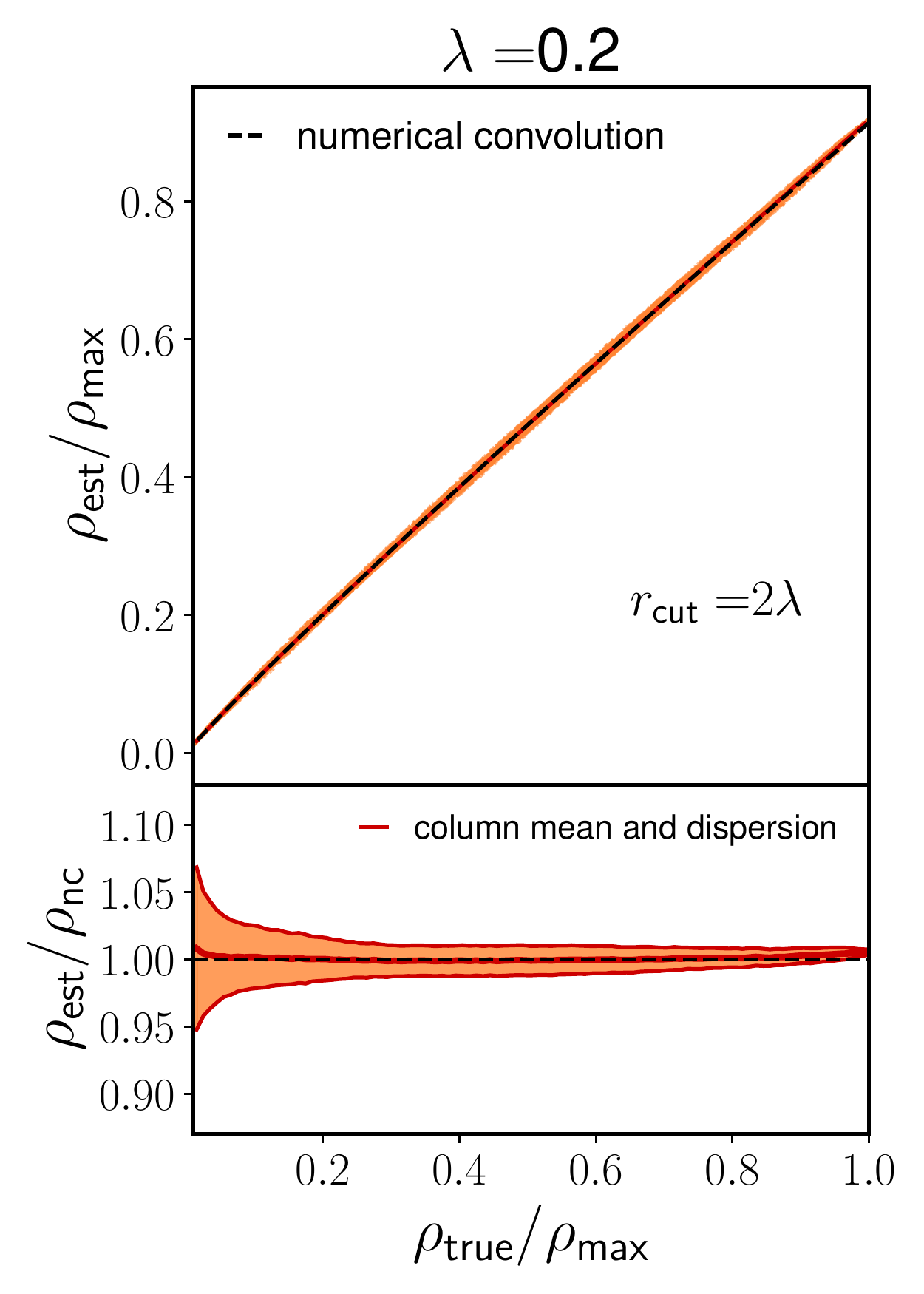}
  \caption{Upper panel: estimated densities for the same spherical point distribution of Section~\ref{sec:sph_dens} plotted 
  against their true values. The densities are estimated using the Gaussian kernel of equation~(\ref{eq:Gaussian_kernel}) 
  with a smoothing length $\lambda= 0.2$ and truncated at a radius $r_{\mathrm{cut}}=2\lambda$ and
  closely follow the convolution of the true density profile with the same kernel, indicated by a black
   dashed line. Lower panel: the ratios of the estimated densities and their expected values. The red lines indicate 
   the column mean and corresponding dispersion.}
  \label{fig:dens-dens}
\end{figure}

\begin{figure}
 \centering
 \includegraphics[width=\columnwidth]{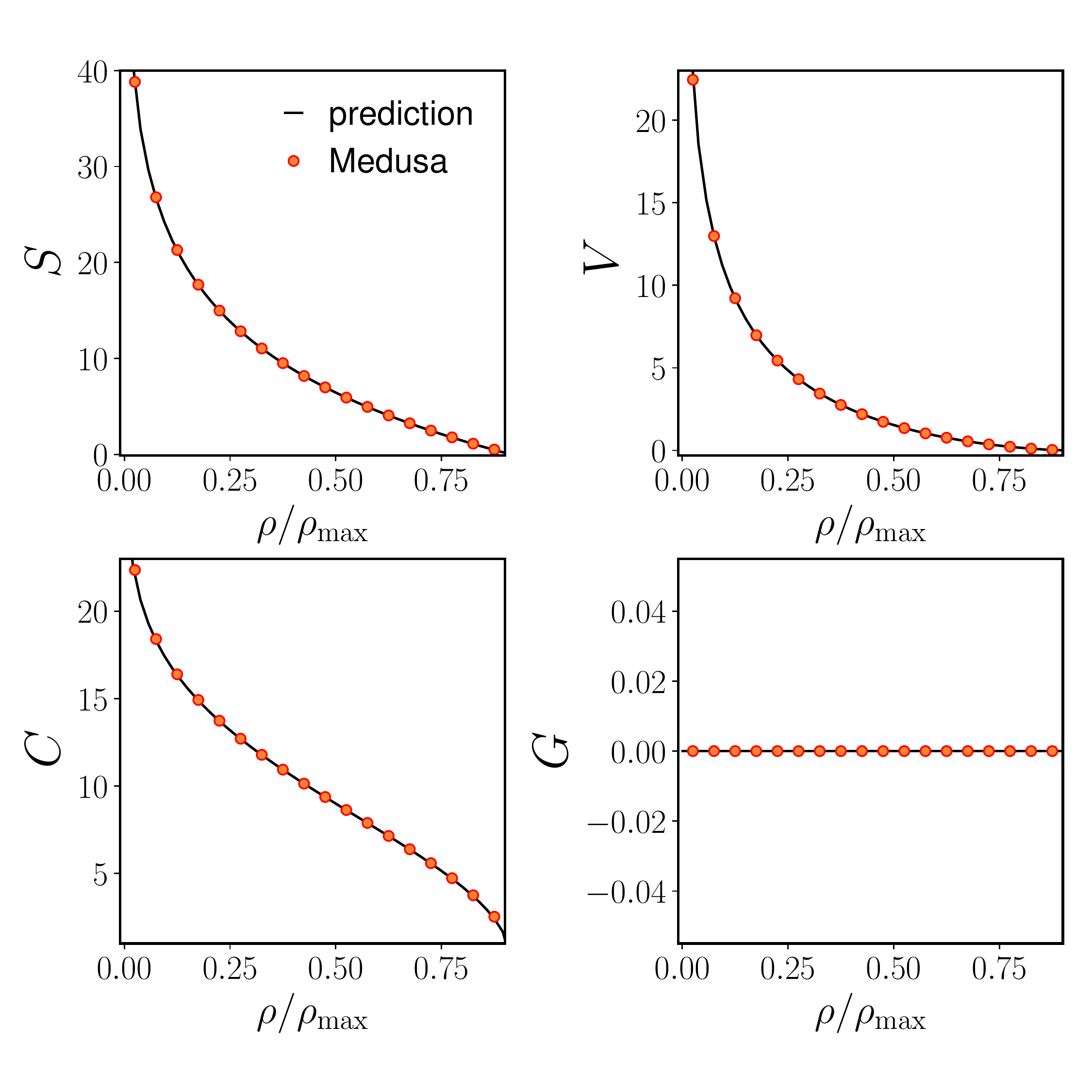}
  \caption{MFs of the  same spherical point distribution of Section~\ref{sec:sph_dens} but inferred from density estimates
  based on a Gaussian kernel with a smoothing length $\lambda= 0.2$ and truncated at a radius $r_{\mathrm{cut}}=2\lambda$, 
  expressed as a function of the normalized density $\rho/\rho_{\mathrm{max}}$. The black lines shows the theoretical 
  predictions corresponding to the convolved density profile.}
  \label{fig:sphere_kernel}
\end{figure}

Fig.~\ref{fig:dens-dens} shows the densities estimated at each point of the spherical Gaussian 
distribution of Section~\ref{sec:sph_dens} by applying a Gaussian kernel with a smoothing length $\lambda=0.2$ 
and a truncation radius of  $r_{\mathrm{cut}}=2\lambda$, which follow closely the true profile, which is indicated 
by a black dashed line. 
Fig.~\ref{fig:sphere_kernel} shows the MFs measured using these density estimates and the corresponding 
theory predictions computed using the convolved density profile. The measurements match the theory 
predictions remarkably well, with a similar level of agreement as for the case in which the 
true densities were used, which was discussed in Section~\ref{sec:sph_dens}. 
Note that, as the convolution with the Gaussian kernel reduces the maximum densities in the profile, 
the highest density threshold considered in this case is $\rho/\rho_{\mathrm{max}}=0.875$. 
We tested the impact of using different values of $\lambda$ and $r_{\rm cut}$ and found 
similar results but with a larger variance. 

When this method is applied to realistic point distributions such as galaxy catalogues, the smoothing length 
and truncation radius of the kernel need to be adjusted to provide the necessary smoothing to avoid 
discreteness effects without erasing too much information.
In principle, steps (ii) and (iii) of the \medusa code  described in Section~\ref{sec:medusa-code}
could be applied to density estimates obtained using a different approach. 
Other possibilities include non-parametric methods in which the densities are derived from the size of the 
Voronoi or Delaunay cells \citep[e.g.,][]{Schaap2000}. 
Although these approaches can better resolve high-density regions due to their varying resolution, 
we have found that these density estimates are highly affected by Poisson noise in the low density regions 
of sparse samples, and are therefore not optimal for the analysis of real galaxy catalogues. 
An additional advantage of using an isotropic Gaussian kernel with a fixed smoothing length is 
that it is more convenient to compute theory predictions.

\section{Minkowski functionals of the Minerva HOD galaxy catalogues}
\label{sec:mf_hod}
\subsection{Real-space measurements}
\label{sec:mf_hod_real}

\begin{figure*}
 \centering
 \includegraphics[width=0.99\textwidth]{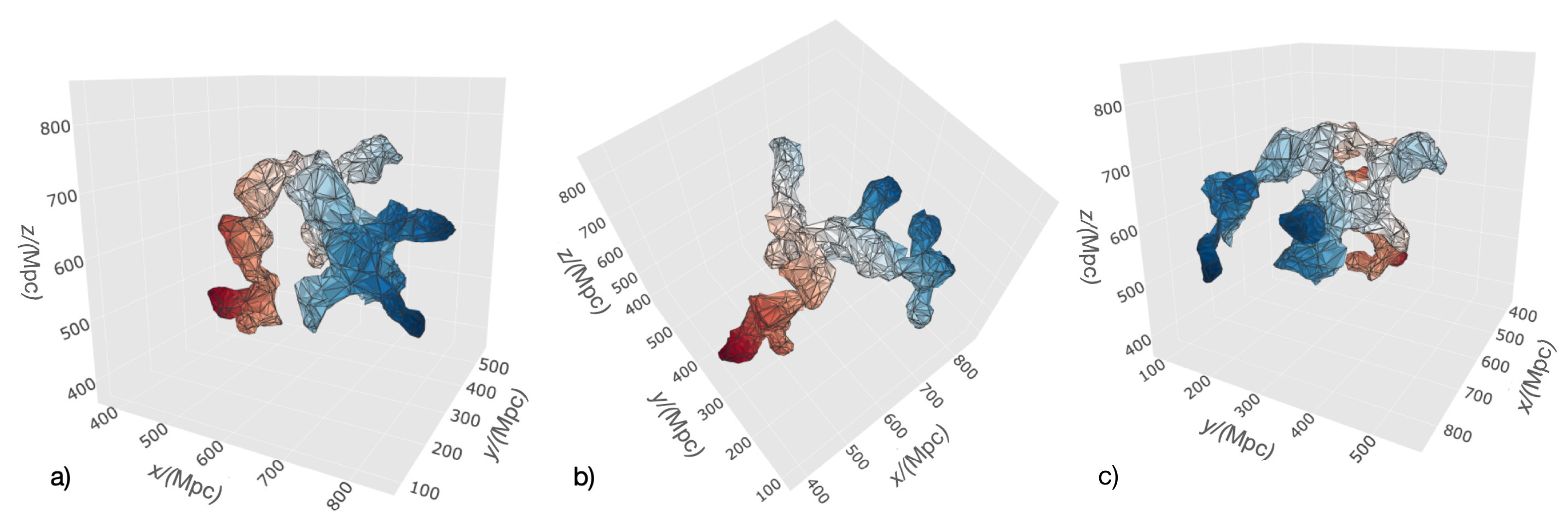}
  \caption{Isodensity surface of one structure found in the first Minerva HOD galaxy catalogue at a density threshold of $\delta = 0.584$.}
   \label{fig:hod_isosurface}
\end{figure*}

\begin{figure}
 \centering
 \includegraphics[width=\columnwidth]{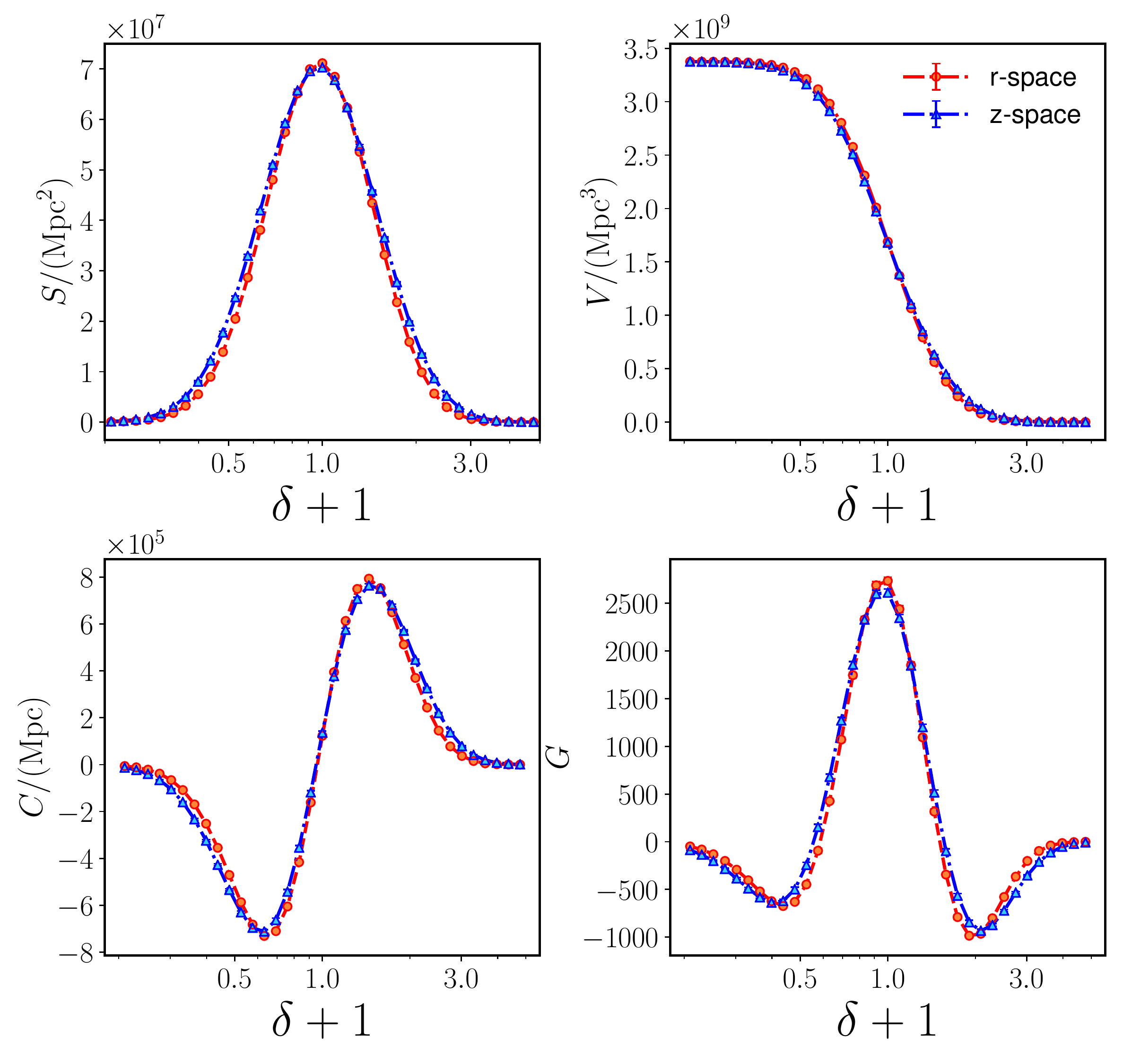}
  \caption{Mean MFs of the 300 Minerva HOD catalogues as a function of the density 
  contrast $\delta$ for the real- and redshift-space galaxy density fields (orange and blue, respectively). The error bars corresponding to the standard deviation from the 300 realizations are of the size of the points or smaller and therefore not visible.
  The densities were estimated with a smoothing length $\lambda$ corresponding to the mean interparticle separation and a truncation radius $r_{\mathrm{cut}}=3\lambda$.}
   \label{fig:mf_zs}
\end{figure}

After validating the performance of \medusa  for several test cases with different geometries and topologies, we now 
show the results obtained by applying the code to synthetic cosmological galaxy samples.  We use catalogues derived from 
the set of 300 N-body simulations Minerva  \citep{2016MNRAS.457.1577G, Lippich2019}. These simulations represent independent realizations of the 
same cosmology, corresponding to the best-fitting flat \textLambda CDM model from the WMAP+BOSS DR9 analysis
of \citet{2013MNRAS.433.1202S}. They are characterized by a total matter and baryon densities 
$\Omega_{\rm m} = 0.285$ and $\Omega_{\rm b} = 0.046$, a Hubble constant of $H_0 = 69.5\,{\rm km}\,{\rm s}^{-1}{\rm Mpc}^{-1}$, a 
scalar spectral index $n_{\rm s}= 0.968$, and a linear-theory rms mass fluctuation in spheres of radius $12\,{\rm Mpc}$, 
$\sigma_{12}=0.805$ \citep{Sanchez2020}.
The evolution of the dark matter density field was simulated with $1000^3$ dark matter particles per realization in a cubic box of 
side length $L = 1.5 h^{-1}$Gpc with periodic boundary conditions. Halos were identified with a standard Friends-of-Friends 
(FoF) algorithm. To create a synthetic galaxy catalogue, the halos  of the snapshot at $z=0.57$ were populated using the halo occupation 
distribution (HOD) parametrization of  \citet{2007ApJ...667..760Z}. This redshift corresponds to the mean 
redshift of the CMASS sample of BOSS and \citet{2016MNRAS.457.1577G} chose the HOD parameters such 
that the monopole of the mean correlation function from the resulting sample matches the 
one measured from the CMASS galaxies.

As a first step, we estimated the number density, $n_{\mathrm{est}}$, at the position of each galaxy by smoothing the distribution 
with a Gaussian kernel as described in Section\,\ref{sec:densest}. We used a smoothing length corresponding to the mean interparticle 
separation, $\lambda =19.7$\,Mpc, close to what was found to be the optimal smoothing length for 
BAO reconstruction in the final BOSS analyses \citep{2017MNRAS.470.2617A}. This smoothing length is sufficiently 
large to avoid 
discreteness effects, but without erasing too much information on small scales. The truncation radius of the kernel was set to  
$r_{\mathrm{cut}} = 3\lambda$, which gives the highest signal-to-noise. The density contrast at the position of each galaxy 
was obtained as $\delta = n_{\mathrm{est}}/\bar{n} - 1$, where $\bar{n}$ is the mean number density.

We computed the MFs on 35 density thresholds equispaced in logarithmic scale around the mean density contrast 
$\delta = 0$. Fig.\,\ref{fig:hod_isosurface} shows a section of the isodensity surface corresponding to the 
threshold $\delta_{\rm th} = 0.584$ viewed from three different angles. This sample is sparser than the test samples
of Section~\ref{sec:test_samples}, which makes the triangles contributing to the surface more visible than for the toroidal 
profiles of Fig. \ref{fig:torus_surface}. This structure 
has a hole in the center that is visible in panel c), and can then be described by a local genus of 1.

Fig.\,\ref{fig:mf_zs} shows the mean global Minkowski functionals from the 300 Minerva realizations as a function 
of $\delta_{\rm th}$. In logarithmic scale, the shape of the MFs resembles that of the Gaussian predictions from 
Fig.\,\ref{fig:gauss}, but the genus is clearly not symmetric and exhibits different depths for the two minima. 

\begin{figure}
 \centering
 \includegraphics[width=\columnwidth]{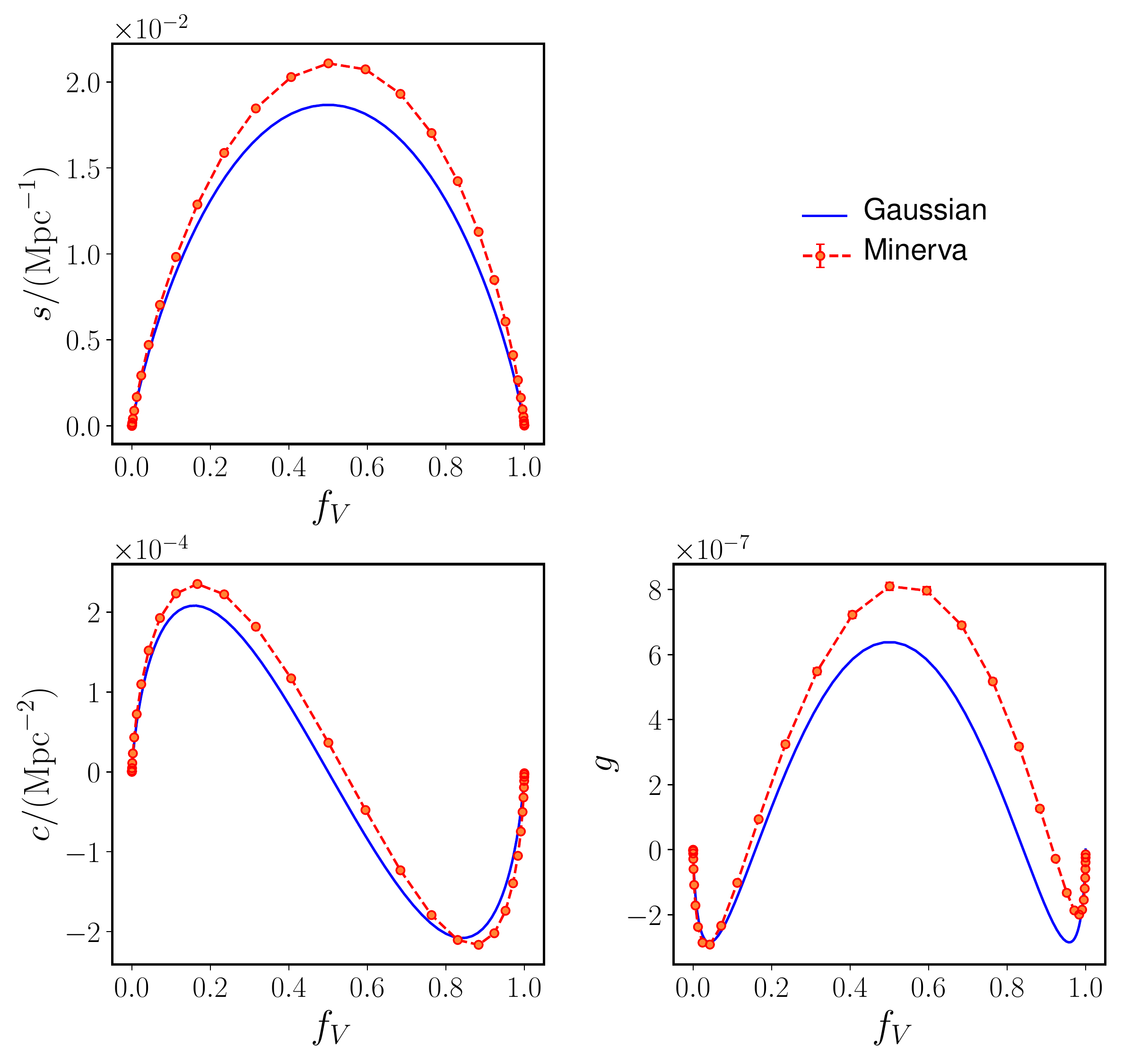}
  \caption{MF densities for the mean of the 300 Minerva HOD catalogues measured from the same smoothed galaxy 
  density field in real space as in Fig.\,\ref{fig:mf_zs}, but plotted as a function of the volume-filling fraction $f_V$.}
  \label{fig:mf_rs_ffv}
\end{figure}

In order to compare MF measurements to theory predictions, the MF densities are typically expressed as functions of the volume-filling fraction $f_V$. The advantage of this is that the MF densities are expected to be invariant under any local monotonic transformation, if the threshold is adjusted such that that it gives the same volume-filling fraction \citep{2013MNRAS.435..531C}.
Fig.\,\ref{fig:mf_rs_ffv} shows the measured mean Minkowski functional densities and the corresponding Gaussian predictions as a function of $f_V$. The Gaussian predictions are obtained from 
equations~(\ref{eq:gauss-vol})-(\ref{eq:gauss-gen}) using the measured mean galaxy power spectrum multiplied 
by the Fourier transform of the smoothing kernel. There are clear differences between the measurements and 
the Gaussian predicitions. 
In particular, the asymmetry of the genus is also obvious here, with the Gaussian prediction 
providing a better match to the measurement at low $f_V$ values. The measured power spectrum, which is well in 
the non-linear regime, is dominated by the high-density regions. Hence, it is to be expected that the 
Gaussian prediction derived from it is in better agreement with the measurements at the high-density end, 
which corresponds to low $f_V$ values.

It is clear that the Gaussian model cannot be used to analyse the MFs of galaxy 
catalogues with comparable number density and redshift as our HOD sample. Since the measurements of the 
surface area, curvature and, in particular, the genus are sensitive to the non-Gaussian features of the density field, 
they contain complementary information to that of the galaxy power spectrum. We will explore the cosmological 
information content of these measurements in detail in upcoming work. In the next sections, we will focus on 
two important observational effects that must be taken into account before measuring the MFs of a 
real galaxy survey, namely RSD and AP distortions.

\subsection{The effect of redshift-space distortions}
\label{sec:rsd}

\begin{figure}
 \centering
 \includegraphics[width=\columnwidth]{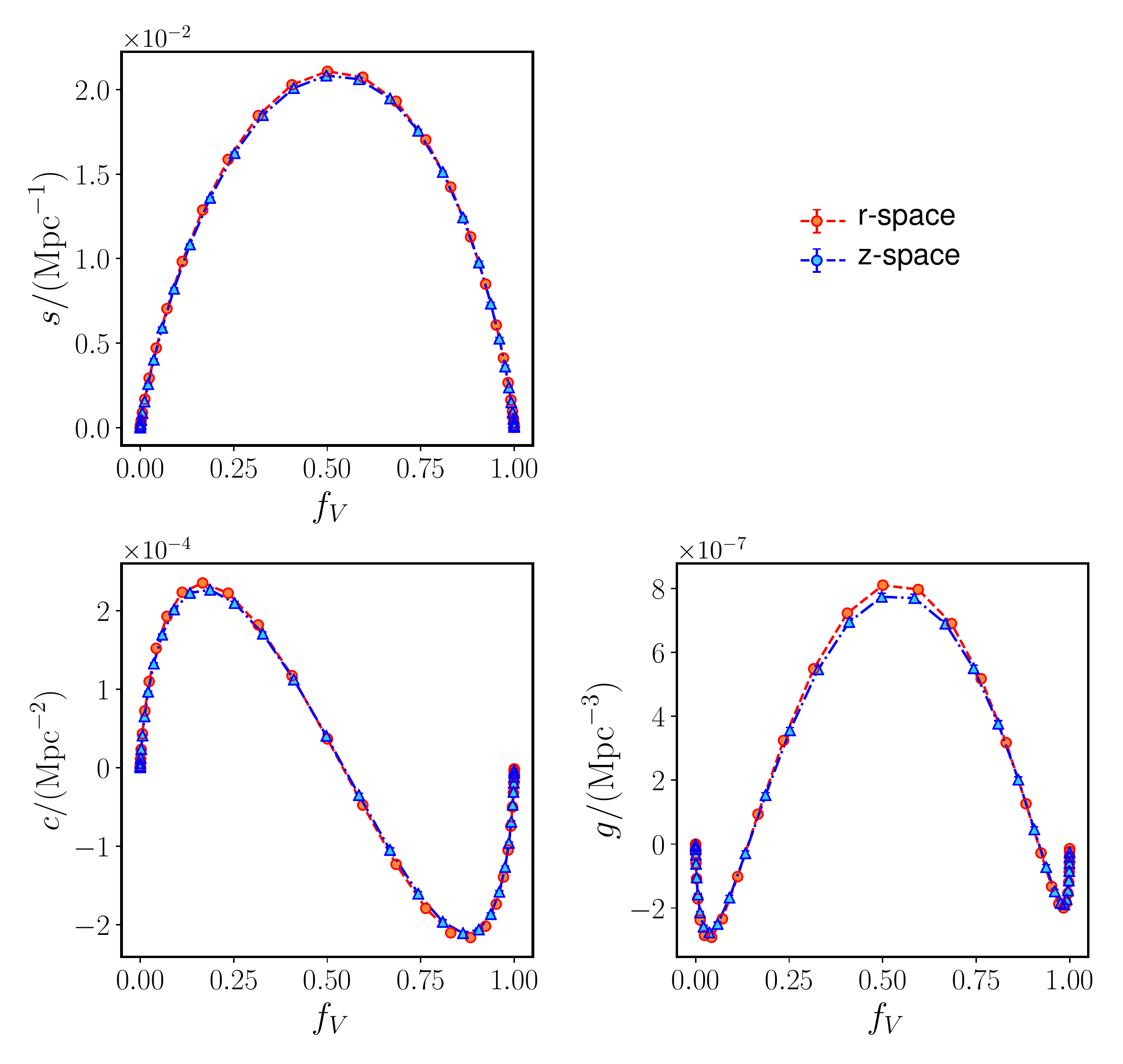}
  \caption{MF densities for the mean of the 300 Minerva HOD catalogues measured from the same smoothed galaxy 
  density field in real and redshift space as in Fig.\,\ref{fig:mf_zs}, but plotted as a function of the volume-filling fraction $f_V$.}
  \label{fig:mf_zs_ffv}
\end{figure}

To study the effect of RSD on the MFs, 
 we distorted the positions of the HOD galaxies by taking into account the 
component of their peculiar velocities along one Cartesian axis of the box, which 
was treated as the line-of-sight direction.  Since the total volume 
and number density are not altered by RSD, we used the same smoothing length as in 
Section\,\ref{sec:mf_hod_real} to estimate the densities at the distorted galaxy positions.
Fig.\,\ref{fig:mf_zs} compares the measurements of the MFs in real and redshift space. 
The amplitudes of both sets of measurements are very similar, but 
the redshift-space MFs appear to be stretched towards lower and 
higher densities than $\delta = 0$ compared to the corresponding ones in real-space.

Fig.\,\ref{fig:mf_zs_ffv} shows the same measurements from Fig.\,\ref{fig:mf_zs}, but plotted as 
functions of the volume-filling fraction, $f_V$. 
Expressed in this way, the agreement between the MFs in real and redshift space is significantly improved, 
with only small deviations in their amplitude. 
The surface measurements agree at a 2\% level, while the deviations in the curvature and genus are smaller than 
5\% (except for the density thresholds where these MFs are close to zero). RSD do not correspond to 
a monotonic transformation of the density field. Nonetheless, on average, the mapping from the real-space 
density threshold, $\delta_{\mathrm{rs}}$, to the corresponding value in redshift space, 
$\delta_{\rm zs }(\delta_{\rm rs})$, can be well described by matching the values of $f_V$ in the two 
spaces (although the scatter for the individual densities is large). 
For this reason, the global effect of RSD on the MFs of the Minerva HOD galaxy catalogues is small 
when these are expressed as functions of $f_V$. 
However, this result cannot be generalised to other samples with different number densities 
or mean redshifts without careful study.

The fact that RSD have only a small effect on the MF densities when expressed in terms of $f_V$
implies that it should be possible to probe the impact of deviations from Gaussianity or the 
sensitivity to the underlying cosmology without a detailed characterization of the 
mapping between real and redshift space. However, an accurate 
modelling of such mapping would open up the possibility to use the measurements of all four 
MFs and to extract constraints on the growth-rate of cosmic structure. We leave such 
analysis for a future study.

\begin{figure}
 \centering
 \includegraphics[width=\columnwidth]{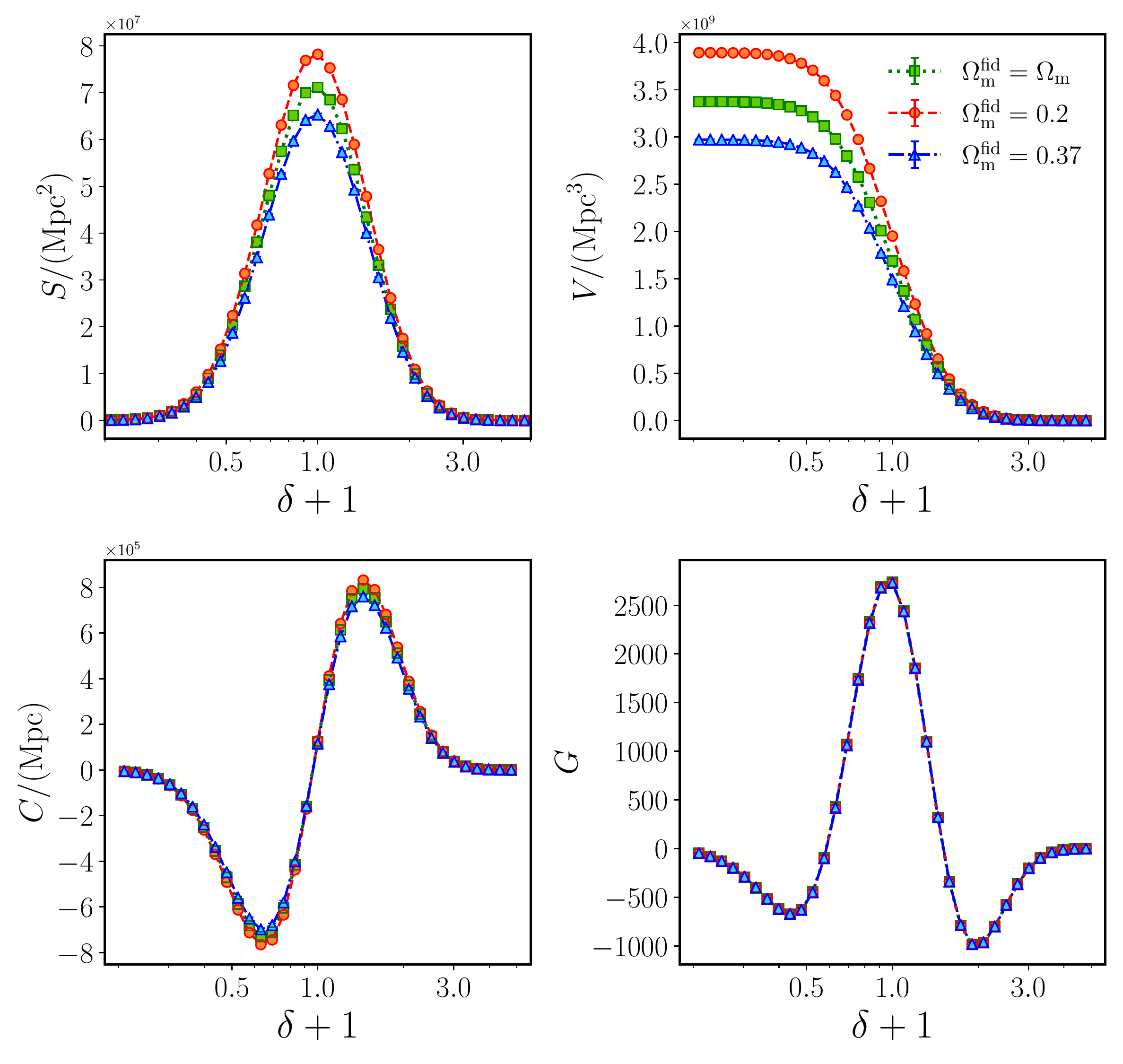}
  \caption{Minkowski Functionals for the mean of the 300 Minerva HOD catalogues as a function of the galaxy density contrast 
  $\delta$ for three AP distorted boxes with different fiducial $\Omega_{\rm m}'$ : a) the undistorted box 
  with $\Omega_{\rm m} ' = \Omega_{\rm m}$, b) $\Omega_{\rm m} ' = 0.20$, b) $\Omega_{\rm m}' = 0.37$ .The densities 
  are estimated with a kernel of a smoothing length $\lambda$ corresponding to specific the interparticle and a truncation radius $r_{\mathrm{cut}}=3\lambda$.}
   \label{fig:mf_ap}
\end{figure}

\subsection{The effect of Alcock-Paczynski distortions}
\label{sec:apd}
The MFs measured from a real galaxy survey will depend on the 
fiducial cosmology assumed to transform the observed redshifts 
into comoving distances. Any difference between this cosmology and the 
true underlying one gives rise to AP distortions \citep{Alcock1979}. 
The modelling of AP distortions is standard in the analysis of two-point statistics, 
but has mostly been ignored for MFs. We mimic the effect of AP distortions on our 
Minerva HOD samples by distorting the galaxy positions by
\begin{equation}
  x_{\perp}' = \frac{D_{\rm M}'(z)}{D_{\rm M}(z)} x_{\perp}\equiv q_{\perp}^{-1} x_{\perp},
  \label{eq:xperp}
\end{equation}
 for the two Cartesian axes perpendicular to the line of sight and
\begin{equation}
  x_{\parallel}'= \frac{H(z)}{H'(z)}x_{\parallel} \equiv q_{\parallel}^{-1}x_{\parallel},
  \label{eq:xpara}
\end{equation} 
 for the line-of-sight coordinate. We computed the values of the comoving angular-diameter distance, $D_{\rm M}(z)$, and the 
 Hubble parameter $H(z)$ using the true underlying matter density of the Minerva simulations and 
 the fiducial values, $D_{\rm M}'(z)$ and $H(z)'$, using two different fiducial matter densities 
 $\Omega_{\rm m} ' = 0.20$ and $\Omega_{\rm m} ' = 0.37$. 
We applied the same smoothing procedure as in Section\,\ref{sec:mf_hod_real} to the AP distorted HOD 
galaxy samples, where we again set the smoothing scale $\lambda$ as the mean inter-particle separation
and $r_{\rm cut}=3\lambda$. 
As the volumes of the AP distorted boxes change with respect to the undistorted reference one, also the 
mean interparticle separations, and therefore the corresponding values of $\lambda$ and $r_{\rm cut}$, 
are adjusted accordingly. 

We used \medusa to measure the MFs of the resulting density fields using the same 
density thresholds as in Section\,\ref{sec:mf_hod_real}. Fig.\,\ref{fig:mf_ap} shows the mean global MFs 
as function of the density contrast $\delta$ of the original boxes (green points) and the two distorted 
cases (orange and blue points). There are obvious differences in the amplitudes of $S$, $V$, and $C$ 
for the three different choices of fiducial matter densities. As the topology of the galaxy density field 
is not changed by the coordinate transformations of equations~(\ref{eq:xperp}) and (\ref{eq:xpara})
the genus is the same in all cases. 

\begin{figure}
 \centering
 \includegraphics[width=\columnwidth]{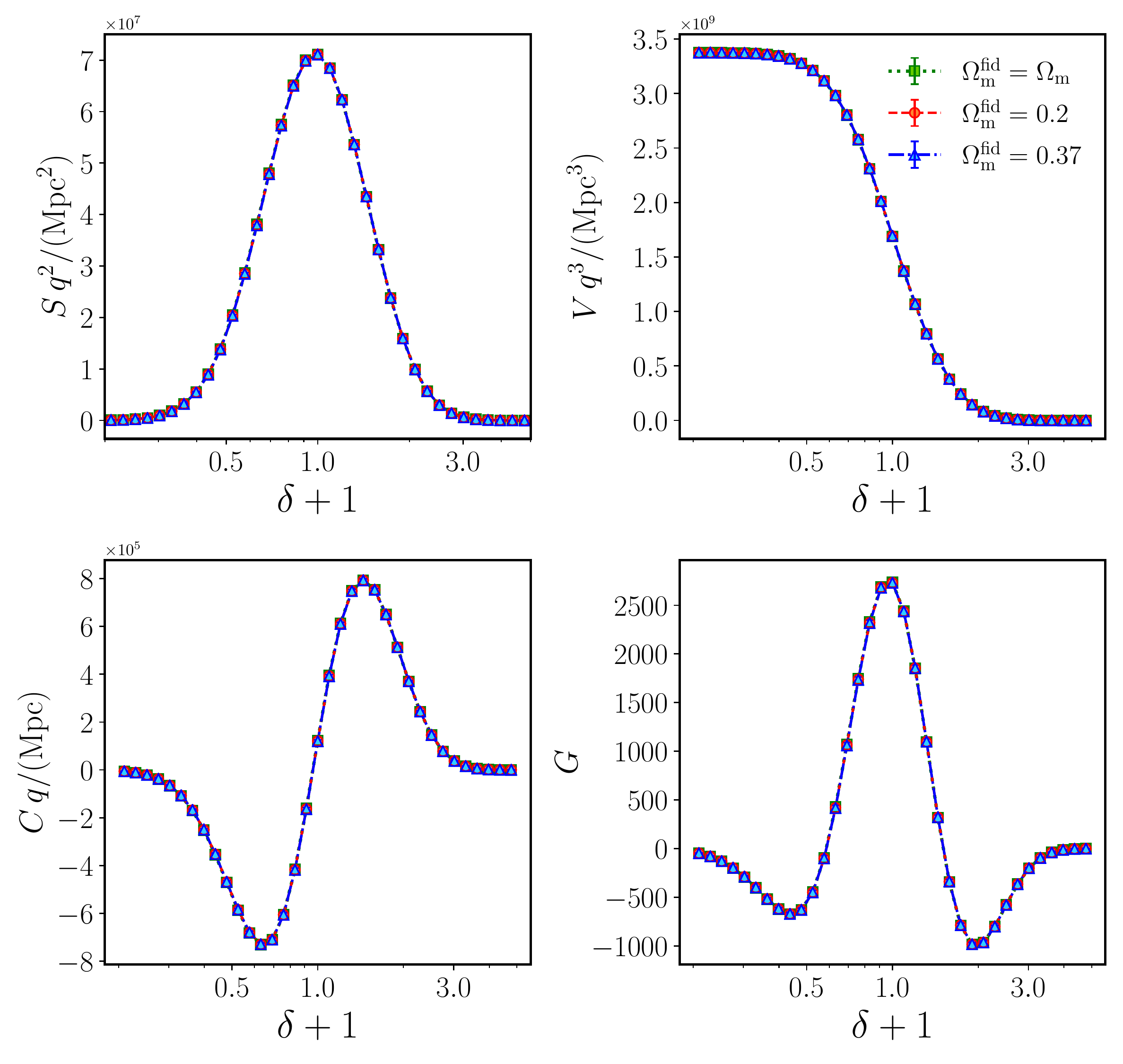}
  \caption{The same as Fig.\,\ref{fig:mf_ap}, but now the MFs are normalized by the corresponding 
  powers of the isotropic AP parameter $q$.}
   \label{fig:mf_ap_q}
\end{figure}

As MFs are angle-averaged measurements, they are sensitive to the isotropic AP parameter,
   \begin{equation}
   q = \left(q_{\perp}^2q_{\parallel}\right)^{1/3}=\frac{D_{\rm V}(z)}{D_{\rm V}'(z)},
  \end{equation}
which depends on the  volume-averaged distance 
\begin{equation}
D_{\rm V}(z) = \left(D_{\rm M}(z)^2\,cz/H(z)\right)^{1/3}.
\end{equation}
The coordinate transformation associated with AP distortions is described by the Jacobian 
of the volume and surface integrals of the MFs. Following from this, the global MFs transform 
under AP distortions as
  \begin{align}
  S &= q^2\,S',  \label{eq:mf_qs} \\ 
  V &= q^3\,V',   \label{eq:mf_qv} \\
  C &= q\,C',   \label{eq:mf_qc}
  \end{align}
  while the genus remains unaffected. Equivalently, the AP distorted MF densities can be rescaled by 
  the factors $q^\alpha$, 
  with $\alpha=0$, -1, -2, -3 for $f'_V$, $s'$, $c'$, and $g'$ to obtain the undistorted MF densities.
Fig.~\ref{fig:mf_ap_q} shows the global MFs rescaled by the corresponding powers of $q$, 
which are in excellent agreement with the 
undistorted reference measurements. 

The correction factors of equations\,(\ref{eq:mf_qs}) -- (\ref{eq:mf_qc}) must be taken into account 
before any model of 
the MFs can be compared against measurements inferred from galaxy redshift surveys. 
They also show that these measurements can be used to constrain $q$, and hence the volume-averaged 
distance $D_{\rm V}(z)$. This was the approach followed by 
\citet{2014MNRAS.437.2488B}, who used Gaussian theory predictions to derive constraints on 
$D_{\rm V}(z)$ from the differential MFs of WiggleZ. 
As we discussed in Section~\ref{sec:mf_hod_real}, the Gaussian predictions do not give a correct description of 
the MFs of our HOD catalogues, indicating that the derivation of unbiased constraints on $D_{\rm V}(z)$ 
from real galaxy samples with similar clustering properties (such as the BOSS CMASS sample) would require 
a more accurate treatment of the impact of non-linearities on the MFs.

\section{Discussion and Conclusions}
\label{sec:conclusions}

We presented {\sc MEDUSA}, an implementation of a robust and accurate method to estimate the MFs
of a three-dimensional distribution of points with known densities. 
Given a density threshold, \medusa performs two main steps: First, closed triangulated 
surfaces are constructed from the tetrahedra of the Delaunay tessellation of the points. 
This is done by selecting the tetrahedra that contain vertices with densities above and 
below the threshold. The isodensity surface is then defined by linear interpolation 
of the densities of those vertices. 
The second main step is the evaluation of the four MFs, volume, 
 surface area, integrated mean curvature and Euler characteristic, of the triangulated 
 surface by summing over the contributions of all tetrahedra and triangles contributing to it.
An earlier implementation of the same basic algorithm was described in \citet{2004BAAA...47..377Y}. 
 We extended previous works on the estimation of MFs on triangulated surfaces by implementing 
 periodic boundary conditions,  which are essential for the analysis of N-body simulations. 
 The MFs computed by \medusa can be used to study the geometry and topology of 
 the sample being analysed.

We tested \medusa by applying it to estimating the MFs of test samples with different geometrical 
and topological properties. We considered spherical, ellipsoidal and toroidal density distributions
 for which theoretical predictions of the MFs can be computed. In all cases, the estimated volume, 
surface area and extrinsic curvature agree significantly better than one per-cent with the theoretical 
predictions. The Euler characteristic, which is directly related to the genus, is computed exactly.
We also estimated the MFs of different spherical distributions intersecting the edges of a box 
with periodic boundary conditions and found the same level of agreement with the theoretical predictions. 
As a further test sample, we used 100 GRFs with periodic boundary conditions, for which there are 
known analytical predictions of the MFs that are sensitive to the power spectrum of the sample. 
We found a good agreement between the results of \medusa and the theoretical predictions.

After validating the performance of \medusa against test samples with known underlying density 
distributions, we applied it to the analysis of 300 HOD galaxy catalogues from our \textLambda CDM 
Minerva simulations at $z=0.57$. As a first step, we estimated the densities associated with
every point in the sample by means of a Gaussian kernel with a fixed smoothing length matching 
the mean inter-particle separation of the sample. We studied the MFs of the HOD catalogues 
as a function of the density contrast $\delta$ and the volume-filling fraction $f_V$. 
Three important aspects must be taken into account before measurements of the MFs of 
real galaxy redshift surveys can be used as cosmological probes: non-Gaussian signatures due to 
non-linear gravitational evolution, RSD, and AP distortions. Our main conclusions 
on these topics are:
\begin{enumerate}
\item The measured MFs are sensitive to deviations from a Gaussian distribution. In particular 
the asymmetry of the genus is an interesting non-Gaussian signal. Since the MFs encode information 
from high-order statistics, they are promising tools to study the nonlinear galaxy density field as 
complementary probes to the standard two-point analyses.
\item The full measurements of the MFs are affected by RSD. 
An accurate model of the mapping between the real- and redshift-space densities would make it 
possible to extract information on the growth rate of cosmic structure from these measurements. 
However, when expressed as a function $f_V$, the impact of RSD on the MFs of our HOD samples
is greatly reduced. This opens up the possibility to probe deviations from Gaussianity using
redshift-space measurements even without a detailed modelling of RSD.
\item The volume, surface area, and curvature are sensitive to AP distortions. The topology of the density 
field is not affected by AP distortions, and hence the Euler characteristic and genus are independent 
of the fiducial cosmology. The impact of AP distortions on the MFs can be accounted for by rescaling 
the theory predictions by the corresponding powers of the isotropic AP parameter $q$. 
This parameter is directly related to the volume-averaged distance $D_{\rm V}(z)$, which can therefore be constrained
from these measurements.
\end{enumerate}
In forthcoming studies, we aim to analyse in detail the sensitivity of the MFs to specific cosmological 
parameters and to explore the cosmological implications of MFs measurements inferred from real galaxy 
surveys. This will require implementing and testing a description of the MFs of the non-linear galaxy 
density field. The recently published analytic expressions for the  MFs up to second-order corrections in 
non-Gaussianity by \citet{2020arXiv201104954M} and the modelling of RSD by 
\citet{2013MNRAS.435..531C} will be of great help. Our final goal is to advance the 
analysis of MFs as a powerful complement of the standard two-point clustering statistics.

\section*{Acknowledgements}
ML and AGS thank Daniel Farrow, Jiamin Hou, Andrea Pezzotta and Agne Semenaite for their help and 
useful discussions. 

The \medusa measurements and analyses 
presented here have been performed on 
the high-performance 
computing resources of the Max Planck 
Computing and Data Facility (MPCDF) in Garching.

The plots of the 3D isodensity surfaces have been created with the free version of the Python graphing 
library Plotly \citep{plotly}.

\section*{Data Availability}

The data underlying this article will be shared on reasonable request to the corresponding authors.




\bibliographystyle{mnras}
\bibliography{medusa}




\bsp	
\label{lastpage}
\end{document}